\documentclass[twocolumn]{aastex631}

\newcommand{\TESS}{\emph{TESS} }

\newcommand{\purple}{\color{black}}

\received{November 12, 2021}
\revised{\today}

\shorttitle{TIC 5724661: An {\rm sd}B-$\delta$~Sct Binary}
\shortauthors{Jayaraman et al.}
\graphicspath{{./}}

\begin{document}
\title{TIC 5724661: A Long-Period Binary with a Pulsating sdB Star and $\delta$ Scuti Variable}

\correspondingauthor{Rahul Jayaraman}
\email{rjayaram@mit.edu}

\author[0000-0002-7778-3117]{Rahul Jayaraman}
\affiliation{MIT Department of Physics and MIT Kavli Institute for Astrophysics and Space Research, Cambridge, MA 02139, USA}

\author{Saul A. Rappaport}
\affiliation{MIT Department of Physics and MIT Kavli Institute for Astrophysics and Space Research, Cambridge, MA 02139, USA}

\author[0000-0002-6916-8130]{Lorne Nelson}
\affiliation{Department of Physics and Astronomy, Bishop’s University, 2600 College St., Sherbrooke, QC J1M 1Z7, Canada}

\author[0000-0002-1015-3268]{Donald W. Kurtz}
\affiliation{Centre for Space Research, Physics Department, North-West University, Mahikeng 2745, South Africa}
\affiliation{Jeremiah Horrocks Institute, University of Central Lancashire, Preston PR1 2HE, United Kingdom}

\author{George Dufresne}
\affiliation{Department of Physics and Astronomy, Bishop’s University, 2600 College St., Sherbrooke, QC J1M 1Z7, Canada}

\author[0000-0001-7756-1568]{Gerald Handler}
\affiliation{Nicolaus Copernicus Astronomical Center of the Polish Academy of Sciences, Bartycka 18, 00-716 Warsaw, Poland}

\author[0000-0002-1353-4269]{Abdel Senhadji}
\affiliation{Department of Physics and Astronomy, Bishop’s University, 2600 College St., Sherbrooke, QC J1M 1Z7, Canada}
\affiliation{Department of Physics, University of New South Wales Canberra, Northcott Dr., Campbell, ACT 2600, Australia}

\author[0000-0001-9911-7388]{David W. Latham}
\affiliation{Center for Astrophysics | Harvard \& Smithsonian, 60 Garden St., Cambridge, MA 02138, USA}

\author{George Zhou}
\affiliation{Center for Astrophysics | Harvard \& Smithsonian, 60 Garden St., Cambridge, MA 02138, USA}

\author[0000-0001-6637-5401]{Allyson Bieryla}
\affiliation{Center for Astrophysics | Harvard \& Smithsonian, 60 Garden St., Cambridge, MA 02138, USA}

\author{George R. Ricker}
\affiliation{MIT Department of Physics and MIT Kavli Institute for Astrophysics and Space Research, Cambridge, MA 02139, USA}

\begin{abstract}
Using TESS 20-sec cadence data, we have discovered an unusual combination of pulsating stars in what we infer to be a binary system. The binary consists of a standard $\delta$ Scuti star with pulsations over the range 32-41 d$^{-1}$; this is in a likely wide orbit with a hot subdwarf B (sdB) star, which itself has a large-amplitude p-mode pulsation at 524 d$^{-1}$. We establish constraints on the period of the putative binary by using radial velocity measurements of the $\delta$~Scuti star and show that any sdB companion star must orbit with a period greater than a few tens of days. Our identification of this sdB binary serves as an important addition to the relatively small {\purple number} of sdB binaries known to have orbital periods longer than a few days. We model such a binary using {\tt MESA} and find that this system could be formed through stable, nonconservative mass transfer from either a low- or intermediate-mass progenitor, without undergoing a common envelope phase. 
\end{abstract}


\section{Introduction}
\label{sec:intro}

\subsection{Observational properties of hot subdwarfs}

{\purple Hot subdwarf-B stars (sdBs) are core helium-burning stars with thin hydrogen envelopes ($\lesssim 0.01$\,M$_\odot$) that exhibit significant chemical peculiarities.  Hot subdwarf-O stars (sdOs) are even more chemically evolved, with helium burning occurring in a shell around an inert carbon-oxygen (CO) core.  Such stars, which are thought to represent late stages of stellar evolution, are likely derived from the stripped cores of red giants. They usually lie on the blue end of the Extreme Horizontal Branch (EHB) of the Hertzsprung-Russell Diagram (HRD; see, e.g., \citealt{1986ASSL..128...33H,1994ApJ...432..351S, heber}). sdB and sdO stars have surface temperatures of $20,000 \lesssim T_{\rm eff} \lesssim 60,000$ K, with $5 \lesssim \log g \lesssim 6.5$, and masses of $\sim$\,0.47 M$_\odot$.}

We have adopted the following working definitions regarding subdwarfs and their key properties:
\begin{enumerate}
\item sdB stars burn helium (He) in their cores and may also undergo $\alpha$-channel burning of the newly-created carbon in the core (leading to the creation of oxygen). This phase persists for tens of Myr, during which the radius stays roughly constant.
\item sdO stars have a well-defined CO core, with helium burning occurring in a shell around this core, which has completed carbon burning and become inert. Simultaneous hydrogen burning occurs in a thin layer near the surface. This phase is typically shorter than the sdB phase (by a factor of {\purple approximately} 2 to 3), and the radius also remains roughly constant during this phase.
\item sdA (subdwarf A) stars, a newly discovered class of subdwarfs, have poorly constrained properties. Their true nature remains uncertain because there may be a variety of processes leading to their formation {\purple {(see, e.g., \citealt{2019ApJ...885...20Y})}}. While the provenance of these stars remains an important open question, they are not of importance for this work.
\end{enumerate}

\subsection{Pulsating subdwarfs}
The first pulsating sdB (sdBV) star, EC\,14026-2647, was discovered by \cite{1997MNRAS.285..640K}, who found a pulsation with a period of 144\,s. Since then, over 100 such pulsating stars have been discovered, {\purple many of them through space-based missions such as Kepler (including K2) and TESS} (see, e.g.,  \citealt{2017MNRAS.466.5020H, 2021MNRAS.507.4178R}). These stars fall into three categories -- rapid (sdBV$_{\rm r}$) pulsators, with $p$ mode oscillations on the order of a few minutes; slow (sdBV$_{\rm s}$) pulsators, with g~mode oscillations on the order of a few hours, and hybrid pulsators, which exhibit both p~and g~mode oscillations.

\subsection{Formation of sdB stars}

While the different classes of sdBV stars are fairly well-defined, the formation of these objects remains somewhat of a mystery. There have been extensive studies {\purple of} the mechanisms via which sdB stars form; see, for instance, \citet{1976ApJ...204..488M}, \citet{1993ApJ...407..649C}, \citet{1993ApJ...409..387D}, \citet{2002MNRAS.336..449H, 2003MNRAS.341..669H}, \citet{2011MNRAS.410..984J}, \citet{2015ApJ...806..178S},  \citet{2019CoSka..49..264V}, \citet{Sen2019}, and \citet{2020A&A...641A.163V}.

\citet{2002MNRAS.336..449H} specifically compared various formation channels leading to the creation of sdB stars. {\purple They concluded that sdBs in tight binaries ($P_{\rm orb}\lesssim$ 10\,d) were likely formed as a result of common envelope (CE) evolution. On the other hand, they showed that wide systems composed of sdBs + WDs ($P_{\rm orb}\gtrsim$ 400\,d) could, in principle, be formed as the result of stable, yet completely non-conservative Roche lobe overflow (RLOF). Finally, they demonstrated the conditions under which the merger of He WDs (the double helium-WD channel) could lead to the ignition of helium, thereby producing sdBs.

For common envelope evolution to produce a short-period sdB,} a red giant whose mass is at least 2 to 3 times greater than that of its companion must overflow its Roche lobe and achieve a sufficiently high mass transfer rate onto its companion \citep{1976ApJ...204..488M}. Such a high accretion rate precludes the companion star from accreting all of the deposited matter, leading to the formation of a common envelope (see, e.g., \citealt{1941ApJ....93..133K}). The rapid shrinking of the giant's Roche lobe as it loses mass causes dynamically unstable mass transfer, forcing the accreting companion to begin to spiral inside the giant's envelope (see, e.g., \citealt{1976ApJ...209..829W}). If the change in the orbital energy is sufficient to unbind the envelope, then the giant's envelope can be expelled from the binary system on the order of hundreds of years \citep{2017A&A...599A..54X}. If a merger can be avoided, the companion emerges in a tight, circular orbit (periods of hours to days) around the stripped core of the red giant. This stripped core can then evolve onto the EHB and become {\purple an sdB (and/or sdO).} The relatively high proportion of sdB stars observed in short-period binaries suggests that this evolutionary scenario is the most common \citep{2003MNRAS.341..669H}, but there exists an observational bias that favors the discovery of such systems.

{\purple Models of longer-period binaries containing sdBs can also be produced by assuming that the primordial binary, consisting of two main sequence (MS) stars undergoes stable, but (partially) non-conservative, mass transfer, {\purple in which both mass and angular momentum leave the system.} Using a binary population synthesis code, \citet{2003MNRAS.341..669H} concluded that it is possible to produce sdBs in binaries with $P_{\rm orb}\approx$ 100\,d under these assumptions. They refer to this as the ``first stable RLOF channel.'' It should be noted, however, that the evolution of the accretor does not seem to have been computed contemporaneously with that of the donor. This could possibly result in the accretor filling its Roche lobe before the donor has had a chance to evolve to (or completely through) the sdB phase. \citet{2003MNRAS.341..669H} also considered the formation of wide sdB+WD binaries with periods on the order of 1000\,d (``second CE ejection channel''), but their simulations failed to produce any, due to the need for massive WDs in such systems. 

Very recently, \citet{2020A&A...641A.163V} showed that the observed population of wide sdB binaries ($P_{\rm orb}\gtrsim$ 1000\,d) could be robustly reproduced under the assumption that the low-mass primordial primary star (donor) is close to the tip of the red giant branch (helium flash) when rapid, yet stable, non-conservative mass loss occurs as a result of Roche lobe overflow.  This can result in the formation of wide binaries containing sdBs.  Using population synthesis techniques, they also investigated the effects of metallicity. For solar metallicities, they found $P_{\rm orb}\gtrsim$ 1000\,d; for lower metallicity stars they showed that sdB binaries were likely to have $P_{\rm orb}\approx$ 1000\,d.

It is also possible to form sdB/O binary stars via stable, (partially) non-conservative mass transfer in progenitor binaries composed of intermediate-mass, main-sequence stars.} While we know that mass transfer can be partially  non-conservative based on an analysis of Algol-related binaries \citep{2000NewAR..44..111E}, we do not have a good constraint on systemic mass loss (i.e., the fraction of mass ejected from the binary). An extensive grid of more than 3000 progenitor models was calculated by \citet{Sen2019} of potential progenitors of hot subdwarfs, under the assumption of partially non-conservative, stable mass transfer {\purple (and solar metallicity). The primaries of the primordial binaries were chosen to have masses between 1 and 8 M$_\odot$ (with the secondaries having masses of 25\%, 50\%, 80\% and 90\% of the primary), and initial orbital periods of 1--200\,d. Mass transfer was also parameterized so as to be arbitrarily non-conservative (i.e., ranging from 0 to 100\%). That work suggests that sdB-containing binaries could have orbital periods in the range of 10\,d $\lesssim P_{\rm orb} \lesssim 100$ \,d. Thus, they naturally bridge the gap in $P_{\rm orb}$ between sdBs formed as a result of common envelope evolution and those formed from low-mass red giants that undergo rapid mass transfer near the tip of the red giant branch.}

\subsection{TIC 5724661}
As part of its goal to enable precision asteroseismology, the Transiting Exoplanet Survey Satellite (TESS; \citealt{2015JATIS...1a4003R}) has been observing a subset of stars using a novel 20-s cadence since Sector 27, which occurred during 2020 July. This mode can probe frequencies up to a Nyquist limit of 2160\,d$^{-1}$, corresponding to periods as short as 40\,s. TESS short-cadence observations are invaluable in the detection and characterization of new subdwarfs, especially pulsators (see, e.g., Section 6 of \citealt{2021FrASS...8...19L}, and references therein). 

TIC~5724661 was chosen to be observed at 20\,s cadence because it was a known A~star in the instability strip on the HRD that might exhibit high-frequency pulsations. Observations of this star and our subsequent analyses revealed two sets of pulsational frequencies -- one in the typical $\delta$ Scuti frequency range and two other, unexpected, peaks at 524\,d$^{-1}$ and 580\,d$^{-1}$. These two are in the characteristic pulsational frequency range of hot compact stars, like white dwarfs and sdBV$_{\rm r}$ stars. 

In this paper, we first analyze the frequencies of the $\delta$ Scuti pulsations and discuss the modes they represent. Then, we show that the spectrum of TIC~5724661 does not exhibit the chemical abnormalities of a roAp star. Moreover, no significant variations in radial velocity are found over a series of unequally-spaced observations, suggesting a long-period ($\gtrsim$~70 d) orbit. We next discuss the strong evidence for a hot compact companion suggested by the excess ultraviolet (UV) flux in the Spectral Energy Distribution (SED). We then explore possible formation pathways for such a system and contextualize our discovery of a novel system. Note, we will hereafter refer to the A-star component of the binary as the ``secondary'', and the sdB component as the ``primary'', for reasons regarding the evolution of the binary that will be more thoroughly explained in Section \ref{sec:stel-evol}.

\section{Observational data}
\label{sec:obs}
\subsection{TESS Observations of TIC 5724661}
\label{sec:tess}

\begin{table}
\centering
\caption{Properties of TIC~5724661}
\begin{tabular}{lc}
\hline
\hline
Parameter & Value   \\
\hline
RA (J2015.5) (h m s)& 23:11:07.84  \\  
Dec (J2015.5) ($^\circ \ ^\prime \ ^{\prime\prime}$) &  -17:13:19.424 \\ 
$T$$^a$ & $11.204 \pm 0.007$ \\  
$G$$^b$ & $11.286 \pm 0.001$  \\ 
$G_{\rm BP}$$^b$ & $11.353 \pm 0.001$  \\ 
$G_{\rm RP}$$^b$ & $11.154 \pm 0.001$  \\ 
$B^a$ & $11.431 \pm 0.129$ \\
$V^a$ & $11.231 \pm 0.010$ \\

$J^c$ & $10.998 \pm 0.020$   \\
$H^c$ & $10.962 \pm 0.024$  \\
$K^c$ & $10.919 \pm 0.023$ \\
W1$^d$ & $10.889 \pm 0.023$ \\
W2$^d$ & $10.917 \pm 0.020$  \\
W3$^d$ & $10.851 \pm 0.162$  \\
W4$^d$ & $> 8.607$  \\

$R$ (${\rm R}_\odot$)$^e$ & $1.32^{+0.09}_{-0.11}$  \\
$L$ (${\rm L}_\odot$)$^e$ & $5.737 \pm 0.8 $  \\

Distance (pc)$^b$ & $ 611 \pm 15$  \\   
$\mu_\alpha$ (mas ~${\rm yr}^{-1}$)$^b$ & $-4.5197 \pm 0.0375$   \\ 
$\mu_\delta$ (mas ~${\rm yr}^{-1}$)$^b$ &  $+4.941 \pm 0.0304$   \\ 
\hline
\hline
\label{tbl:mags}  
\end{tabular}

{\bf Notes.}  (a)\url{exofop.ipac.caltech.edu/tess/index.php}.  (b) {\em Gaia} eDR3 (\citealt{2016A&A...595A...1G}; 
\citealt{2016A&A...595A...2G}; \citealt{2021A&A...649A...1G}).  (c) 2MASS catalog \citep{2006AJ....131.1163S}.  (d) WISE point source catalog \citep{2014yCat.2328....0C}.  (e) This work; see Table \ref{table:mcmc-mle} for details regarding the radius estimate, as well as Table \ref{table:rvs} for RV data.
\end{table}

\begin{figure}
\begin{center}
\includegraphics[width=0.98\linewidth,angle=0]{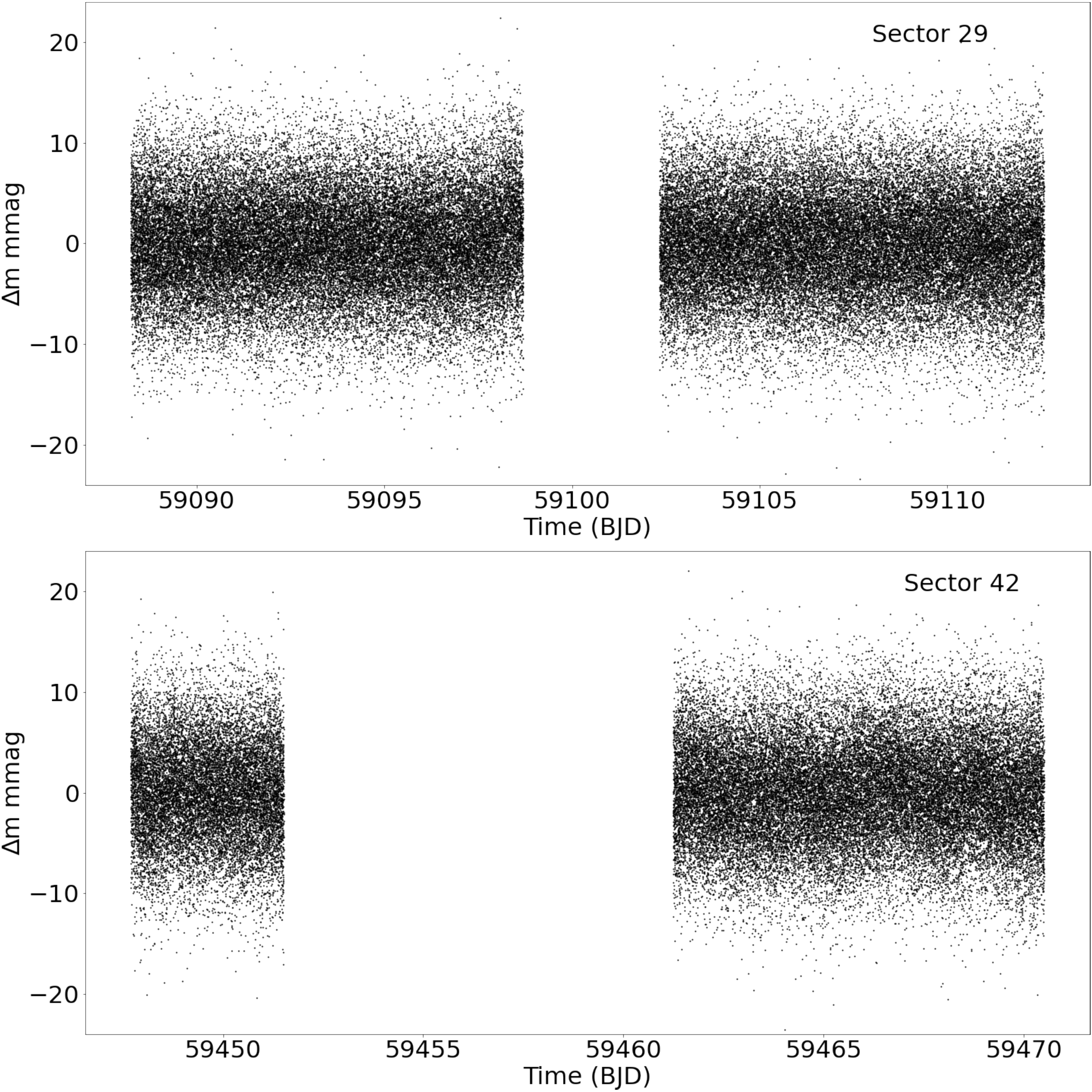}
\caption{The light curve of TIC~5724661 obtained in 20\,s cadence in {\em TESS} Sectors 29 and 42 after processing with the SPOC pipeline \citep{jenkinsSPOC2016}. The pulsations are too rapid and too low in amplitude to discern visually in this compressed figure. Its purposes are (i) to show the two gaps in the data, which affect the spectral window, and (ii) to show the noise level in the 20-s data points. The ordinate scale is Barycentric Julian Date -- 240\,0000.0.}
\label{fig:lc} 
\end{center}
\end{figure}

TIC~5724661 was observed by {\em TESS} in Sector 29 (from 2020 August 26 to September 21) and Sector 42 (from 2021 August 20 to September 16) in both 2\,min and 20\,s cadence. The data are available in both SAP (simple aperture photometry) and PDCSAP (presearch data conditioning SAP) forms. Data processing was done using the Science Processing Operations Center (SPOC) pipeline at NASA Ames \citep{jenkinsSPOC2016}. We used the PDCSAP data from both Sectors 29 and 42 for our analysis after converting the given flux intensity to magnitudes.\footnote{The data from Sector 42 have a large data gap during the first orbit (see Figure \ref{fig:lc}) due to saturation of the CCDs arising from the moon being in the TESS field-of-view, so we rely somewhat less on this dataset.} The Sector 29 data span 24.33\,d with a temporal center point of $t_0 = {\rm BJD}~2459100.41122$, and comprise 88937 data points (after clipping by SPOC to remove outlier points, e.g., those arising from cosmic ray strikes on the detector).\footnote{TESS Sector 29 Data Release Notes: \url{https://archive.stsci.edu/\\missions/tess/doc/tess\_drn/tess\_sector\_29\_drn43\_v02.pdf}}

Figure\,\ref{fig:lc} shows the SPOC-processed light curves with the data gaps between the two orbits making up each \TESS sector. These arise from the lack of observations during data downlink, or saturation of the CCDs due to scattered light from the Earth and the Moon. Such data gaps affect the spectral window, necessitating either analysis with a Discrete Fourier Transform (DFT; see, e.g., \citealt{1985MNRAS.213..773K}) or appropriate corrections, such as re-binning the data into equally-spaced temporal bins.

\subsection{Spectroscopy}
\label{subsec:spec}
We obtained spectroscopic observations of TIC\,5724661 with the Tillinghast Reflector Echelle Spectrograph (TRES, \citealt{Furesz2014}), on the 1.5-m reflector at the Fred Lawrence Whipple Observatory (FLWO) in Arizona, USA. TRES is a high-resolution fiber-fed echelle spectrograph, with a spectral resolving power of $R=44\,000$  over the wavelength region of $3900-9100$\,\AA. A total of {\purple ten} observations were obtained for TIC 5724661 during 2020 December, and between 2021 September and 2021 {\purple December}, with peak signal-to-noise ratios per resolution element of $\sim$30 in the Mg b triplet wavelength region. The spectra were extracted and reduced as per \citet{2010ApJ...720.1118B}, with wavelength solutions derived from bracketing Th-Ar lamp exposures. The observing schedule was designed to be sensitive to a companion with $P_{\rm orb} \lesssim 30$~d.

To derive the spectroscopic broadening profiles and radial velocities from each observation, we performed a least-squares deconvolution (LSD, \citealt{1997MNRAS.291..658D}) of each spectrum against a synthetic non-rotating template; this provided both a value for the radial velocity, along with an uncertainty value. We also conducted a multi-order velocity analysis of the spectra, and derived another set of uncertainties for the radial velocity values. We observed that the multi-order uncertainties were around 50\% greater than the LSD uncertainties {\purple in some cases, and agreed with them in other cases}. Values from both sets of analyses are presented in Table \ref{table:rvs}.

Visual examination of the broadening profiles for a set of lines from the sdB companion remained negative, but the line profiles did show night-to-night variability consistent with typical spectroscopic line variations exhibited by $\delta$ Scuti stars. The broadening profiles were fitted with a model kernel accounting for the rotational, macroturbulent, and instrumental broadening terms, as well as the velocity shift of the spectrum. The comparison to model spectra and further analyses (including the use of a rotating template) are described in Section~\ref{sec:spec}. 

\begin{table}
\centering
\caption{Radial Velocity Measurements of TIC 5724661 from the Tillinghast Reflector Echelle Spectrograph (TRES).} 
\begin{tabular}{cccc}
\hline
\hline
\multicolumn{1}{c}{Observation} & \multicolumn{1}{c}{Radial} &\multicolumn{1}{c}{LSD} & \multicolumn{1}{c}{Multi-Order}
   \\
\multicolumn{1}{c}{Date} & \multicolumn{1}{c}{Velocity} &\multicolumn{1}{c}{Error} & \multicolumn{1}{c}{Error} \\
\multicolumn{1}{c}{(BJD-2400000)} & \multicolumn{1}{c}{(km\,s$^{-1}$)} &\multicolumn{1}{c}{(km\,s$^{-1}$)} & \multicolumn{1}{c}{(km\,s$^{-1}$)} \\
\hline
59190.60806 & $-39.85$ & $\pm\,0.17$ & $\pm\,0.48$\\
59196.58549 & $-39.54$ & $\pm\,0.73$ & $\pm\,0.68$ \\
59199.57533 & $-41.06$ & $\pm\,0.80$ & $\pm\,0.75$\\
59200.59383 & $-40.24$ & $\pm\,0.66$ & $\pm\,0.59$\\
59202.58932 & $-41.05$ & $\pm\,0.69$ & $\pm\,1.13$\\
59484.77567 & $-39.03$ & $\pm\,0.63$ & $\pm\,0.65$\\
59487.72156 & $-38.98$ & $\pm\,0.38$ & $\pm\,0.52$\\
59519.69383 & $-37.51$ & $\pm\,0.47$ & $\pm\,0.71$\\
59566.62276 & $-35.20$ & $\pm\,0.20$ & $\pm\,0.65$\\
59567.60356 & $-35.26$ & $\pm\,0.21$ & $\pm\,0.44$\\
59766.95738 & $-32.19$ & $\pm\,0.39$ & $\pm\,0.56$\\
\hline
\hline
\end{tabular}
\label{table:rvs}
\end{table}

The set of eleven measured radial velocities obtained with TRES is given in Table \ref{table:rvs}.

\section{Frequency analysis}
\label{sec:freq-analysis}

\begin{figure}
\begin{center}
\includegraphics[width=\linewidth,angle=0]{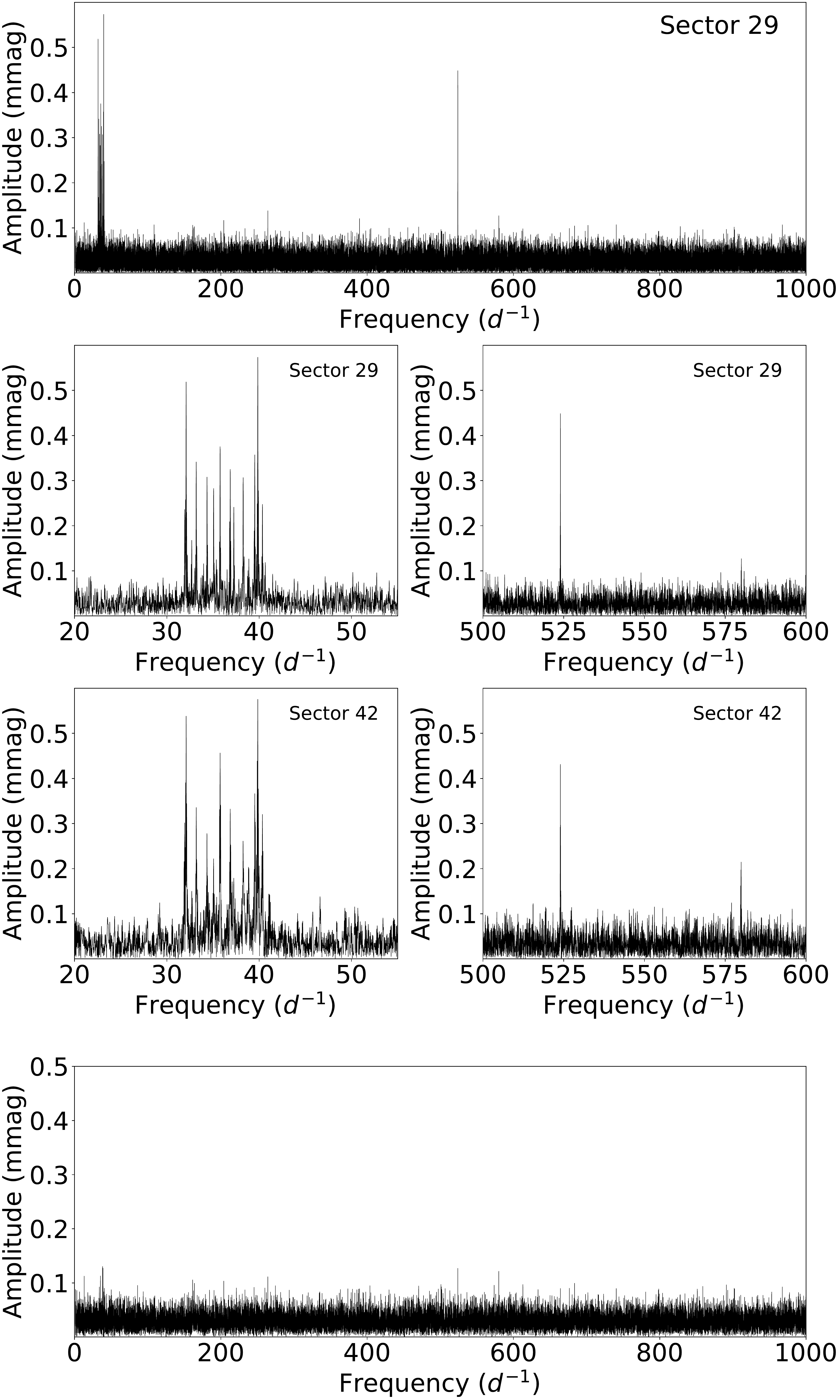}
\caption{The top panel shows the Fourier amplitude spectrum to 1000\,d$^{-1}$ from the Sector~29 light curve; there are no significant peaks between 1000\,d$^{-1}$ and the Nyquist frequency, 2160\,d$^{-1}$. The second panel zooms into a cluster of peaks in the $\delta$~Sct frequency range between $20 - 55$\,d$^{-1}$, along with a single, high-frequency peak at 523.99\,d$^{-1}$ (6.065\,mHz). The third panel shows the $\delta$ Scuti and high-frequency pulsations observed in Sector~42; the peak at 579.85\,d$^{-1}$ (6.711~mHz) increases in prominence between Sectors~29 and 42. The bottom panel shows the Fourier spectrum of the residuals after a non-linear least-squares fit of the 13 highest-amplitude $\delta$~Sct peaks and the peak at 523.99\,d$^{-1}$ (from the sdB star) is subtracted from the data.}
\label{fig:ftd} 
\end{center}
\end{figure}

The {\em TESS} data from both Sectors 29 and 42 were analyzed using a fast Discrete Fourier Transform \citep{1985MNRAS.213..773K} to produce amplitude spectra. The top panel in Fig.\,\ref{fig:ftd} shows the amplitude spectrum out to about half the Nyquist frequency of 2160\,d$^{-1}$, calculated using the Sector 29 data. A cluster of peaks in the $\delta$~Sct frequency range is seen between $26 - 46$\,d$^{-1}$, and a single, high-frequency peak is clearly detected at 523.99\,d$^{-1}$ (6.065\,mHz). These are shown at higher frequency resolution in the two middle panels, with appropriate labels indicating the sector whose light curve was input to the DFT. {\purple There is an additional peak at 579.85 d$^{-1}$ that increases in prominence in Sector 42, lending further credence to our hypothesis of an unseen hot compact pulsator in this system.}

\subsection{{\purple Mode Identification and Asteroseismology}}

We fitted the 13 $\delta$~Sct frequencies and the most prominent sdBV$_{\rm r}$ frequency to the Sector 29 data using a non-linear least-squares algorithm in order to (a) optimize the frequencies, amplitudes and phases, and (b) determine their uncertainties. Those best-fit parameters are provided in Table~\ref{table:freq-fit}. The frequency range is narrow, and the number of excited modes is relatively small for a $\delta$~Sct star. The frequency solution for the $\delta$ Sct modes derived from the sector 42 is consistent to within the observational errors with the one listed in Table~\ref{table:freq-fit}. The bottom panel of Figure \ref{fig:ftd} shows the amplitude spectrum of the residuals after a non-linear least-squares fit of the 14 aforementioned peaks was subtracted from the data. We believe that the highest-frequency peak, at 523.99\,d$^{-1}$ arises from a pulsation mode in an sdBV$_{\rm r}$ star, as discussed later.

A simple zeroth order relation for a pulsator (first derived using a toy model in \citealt{1879AnP...244..157R}) that relates the pulsation period $P$ and mean density $\bar{\rho}$ is:
\begin{equation}
P \sqrt{\frac{\overline\rho}{\overline\rho_\odot}} = Q,
\end{equation}
where $Q$ is a constant for a given pulsation mode, which is defined by this equation. This can be rewritten in terms of observables as follows:
\begin{equation}
\log Q = \log P + \frac{1}{2} \log g + \frac{1}{10} M_{\rm bol} +\log T_{\rm eff} -6.454,
\end{equation}
Here, $P$ is in days and $\log\,g$ is in cgs units. As a first-order estimate, we use the {\em TESS} input catalog (TIC) values of $T_{\rm eff} = 8400$\,K and $\log g = 4.3$ \citep{2019AJ....158..138S} and estimate $M_{\rm bol} = 1.6$\,mag from the {\em Gaia} parallax and $V$ magnitude. Thus, we can calculate the $Q$-values for the $\delta$~Sct frequencies, which enables us to estimate the radial overtone for these frequencies' modes by comparing them with previously-calculated models, such as those in Table 1 of \citet{1979ApJ...227..935S}. Note that the putative sdB companion is significantly fainter in the {\em TESS} and {\em Gaia} passbands (i.e., in the optical -- see Figure \ref{fig:sed-best-fit}), so its contribution to the total absolute magnitude of the system can be neglected here.

For the two highest amplitude modes which span the frequency range of the $\delta$~Sct pulsations, we find $Q = 0.019$ for the 32.0888-d$^{-1}$ frequency and $Q = 0.015$ for the 39.8553-d$^{-1}$ frequency. Comparing these with model 4.4 in \cite{1979ApJ...227..935S} suggests that modes in the $\delta$~Sct star range in radial overtone between $n \sim 2 - 4$. This is a narrow range of overtones, and the number of observed frequencies in the range requires most of the associated modes to be nonradial. At first glance, TIC~5724661 seems to be a relatively hot $\delta$~Sct star, and since hotter stars tend to pulsate in higher radial overtones \citep{1975ApJ...200..343B}, $n \sim 2 - 4$ radial overtones are not unexpected. However, we are also cognizant of the fact that the temperature estimate given in the TIC may be inflated due to an unresolved sdB companion; more details are discussed in Section \ref{sec:sed}.

Three of the peaks are nearly equally spaced in frequency: 34.3520, 35.0681, and 35.7721\,d$^{-1}$. The separations between pairs of these peaks are $0.7161 \pm 0.0025$ and $0.7040 \pm 0.0024$\,d$^{-1}$; these separations themselves differ only by $0.012 \pm 0.003$\,d$^{-1}$. Despite this small difference, the formalism provided in \citet{1992ApJ...394..670D} appears to suggest that this triplet does not arise from rotational splitting. However, if we \textit{do} assume rotational splitting, we can crudely estimate $P_{\rm rot} = 1.4$~d (neglecting the Ledoux rotational splitting constant $C_{n,\ell}$). Moreover, because we know that the binary contains a $\delta$ Scuti star, we can use the illustrative values $\log g = 4.3$ and $M \sim $2\,M$_\odot$ to derive a crude radius estimate of $R \sim 1.6$\,R$_\odot$. These values, along with the rotational period estimate, predict that $v_{\rm eq} = 60$\,km~s$^{-1}$. The spectroscopic estimate of $v\sin\,i$, 39.9 $\pm$ 0.9\, km~s$^{-1}$, suggests that the rotational axis of the star is tilted $\sim$40$^\circ$ with respect to our line of sight (here, we use the convention that 0$^\circ$ is parallel to our line of sight). We emphasize that these are only first-order estimates; further analysis and modeling (discussed in Sections \ref{sec:sed} and \ref{sec:stel-evol}) can better constrain these parameters.

\begin{table}
\centering
\caption{A non-linear least squares fit of 13 $\delta$~Sct frequencies and 1 sdBV frequency to S29 data. The zero point for the phases, $t_0 = 2459100.41122$,  is the center in time of the data.} 
\begin{tabular}{rcr}
\hline
\hline
\multicolumn{1}{c}{frequency} & \multicolumn{1}{c}{amplitude} &   
\multicolumn{1}{c}{phase}  \\
\multicolumn{1}{c}{d$^{-1}$} & \multicolumn{1}{c}{mmag} &   
\multicolumn{1}{c}{radians}   \\
 & \multicolumn{1}{c}{$\pm 0.024$} &   
   \\
\hline
$31.9317  \pm 0.0021 $ & $ 0.246 $ & $ -1.534  \pm 0.102 $ \\ 
 $32.0888  \pm 0.0010 $ & $ 0.522 $ & $ -2.884  \pm 0.047 $ \\ 
 $32.7019  \pm 0.0032 $ & $ 0.160 $ & $ -1.931  \pm 0.151 $ \\ 
 $33.1820  \pm 0.0015 $ & $ 0.341 $ & $ 1.555  \pm 0.071 $ \\ 
 $34.3520  \pm 0.0017 $ & $ 0.300 $ & $ 0.531  \pm 0.081 $ \\ 
 $35.0681  \pm 0.0019 $ & $ 0.267 $ & $ -1.539  \pm 0.091 $ \\ 
 $35.7721  \pm 0.0014 $ & $ 0.362 $ & $ -0.086  \pm 0.067 $ \\ 
 $36.8673  \pm 0.0015 $ & $ 0.330 $ & $ 2.064  \pm 0.073 $ \\ 
 $37.2562  \pm 0.0021 $ & $ 0.243 $ & $ -1.097  \pm 0.100 $ \\ 
 $38.2666  \pm 0.0016 $ & $ 0.314 $ & $ 3.071  \pm 0.077 $ \\ 
 $39.5224  \pm 0.0014 $ & $ 0.356 $ & $ -2.691  \pm 0.068 $ \\ 
 $39.8553  \pm 0.0009 $ & $ 0.585 $ & $ 1.304  \pm 0.042 $ \\ 
 $40.3589  \pm 0.0020 $ & $ 0.258 $ & $ -1.756  \pm 0.094 $ \\ 
 $523.9899  \pm 0.0011 $ & $ 0.449 $ & $ 2.917  \pm 0.054 $ \\ 
\hline
\hline
\end{tabular}
\label{table:freq-fit}
\end{table}

\subsection{{\purple Possible Sources for the 524 d$^{-1}$ Signal}}
\begin{figure*}
    \centering
    \includegraphics[width=\textwidth]{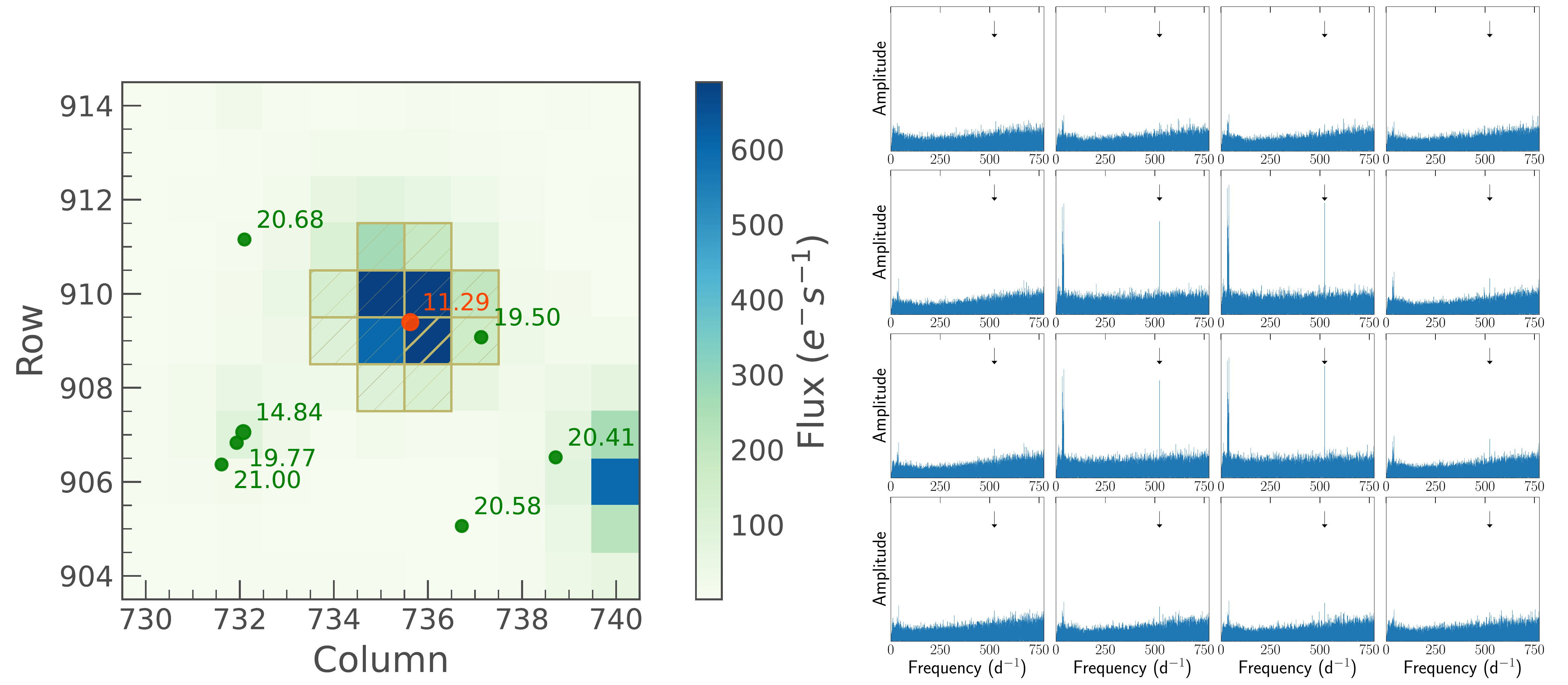}
    \caption{{\purple \textit{Left}: A plot of the optimal aperture for TIC 5724661, marked with shading indicating flux, along with the positions and Gaia magnitudes of nearby stars in the Gaia catalog. It is evident that none of the nearby stars is bright or hot enough to produce a pulsational signal at the observed frequencies. \textit{Right}: A 4$\times$4 subarray of the Fourier transforms of the time series at the pixel level. The result for each pixel has been obtained by convolution with a 3$\times$3 boxcar filter to enhance the statistics, at the cost of decreased spatial resolution (see text for details). The four pixels displaying the highest pulsational signals correspond to exactly the four central pixels of the optimal aperture shown in the left panel. This clearly demonstrates that both the $\delta$~Scuti pulsations and the 524 d$^{-1}$ pulsation arise from within the optimal aperture and that none of the other stars in the TPF causes it. The location of the high-frequency pulsation (from the putative sdB star) is marked with an arrow.}}
    \label{fig:tpf}
\end{figure*}

{\purple Many \TESS~light curves are affected by the blending of targets close to each other on the night sky, in part due to the large size of \TESS~pixels (see, e.g., \citealt{2021ApJS..254...39G}). As a result, we sought to ensure that both the low and high frequency pulsation signals were coming from the same target on the sky. We first downloaded the target pixel file (TPF) for this target and extracted the flux time-series for each pixel. Then, we took the Fourier transform of each individual pixel and produced an 11$\times$11 array of FTs centered on the target star.  Next, we convolved this 11$\times$11 array with a 3$\times$3 boxcar kernel to enhance the statistics, albeit at the cost of some spatial resolution. We found that both the 524 d$^{-1}$ signal and the $\delta$ Scuti pulsations arose from the same region of sky. In particular, they were both strongest when the 3$\times$3 boxcar kernel contained 8 of the 9 pixels from the optimal aperture selected by SPOC to generate the light curve shown in Fig.~\ref{fig:lc}---this aperture is shown in the left panel of Fig.~\ref{fig:tpf}. Outside this region, the amplitude of these pulsations rapidly declined. The right panel of Figure \ref{fig:tpf} displays a 4$\times$4 subarray of the smoothed FTs. The four panels with the highest amplitude peaks at 524 d$^{-1}$ correspond exactly to the four central pixels in the optimal aperture (left panel of Fig.~\ref{fig:tpf}). The left panel, in addition to the optimal aperture, also shows the nearest stars in the Gaia catalog, emphasizing that there are no potential contaminants that could produce a signal of the magnitude we observe.

We obtained a similar result using the newly-developed software tool \texttt{TESS-Localize} \citep{tess-localize}. The likelihood that the 524 d$^{-1}$ signal was indeed coming from TIC\,5724661 and no other contaminating star was found to be $>$99\%, further corroborating our conclusions about the source of the signal.} 

After we confirmed that both signals were coming from the same point on the sky, we explored various possibilities to explain the high-frequency signal. Most A stars with  $v_{\rm eq} \lesssim 100$\,km~s$^{-1}$ are either Am or Ap stars, meaning they exhibit strong metal lines (the distinction arises from the presence of a strong dipole magnetic field in Ap stars; see, e.g., \citealt{2014PhDT.......131M}). So, we would expect TIC~5724661 to show abundance anomalies when examined at high spectral resolution, most probably of the Am kind, as nearly half of A stars near this temperature are Am stars \citep{1973ApJS...25..277S}. However, as discussed in Section \ref{sec:spec}, no abundance anomalies were detectable in our spectra; we may need a data set with a higher spectral resolution to see such anomalies. More evidence against the idea that the 523.99~d$^{-1}$ pulsation arises from a roAp star is the fact that this frequency is over twice the theoretical acoustic cutoff frequency for such a star. None of the observed supercritical roAp pulsations have deviated from this cutoff as strongly (see, e.g., \citealt{2018MNRAS.480.2405H}, and references therein).

Another possible explanation for the high-frequency pulsation observed at 524 d$^{-1}$ is a white dwarf. Many white dwarfs are known to pulsate in this frequency regime, with frequencies associated with g~modes, as opposed to the p~modes in sdBV$_{\rm r}$ stars \citep{2008ARA&A..46..157W}. However, as shown in Figure \ref{fig:ftd}, the amplitude of the high-frequency oscillation is 0.394 mmag. This is 0.036\% of the entire system's light. Using $L = 4\pi\sigma R^2 T^4$, and adopting illustrative values of $0.01~R_\odot$ for the white dwarf radius and $20\,000$\,K for the temperature, we expect the luminosity ratio of the two bodies to be $10^{-3}$, implying the white dwarf pulsates with an amplitude that is $\sim35\%$ of its luminosity. Typical WD pulsation amplitudes are between 1 and 2\% \citep{Winget_1998}; thus, this could not plausibly explain our observations.

Finally, we evaluate the possibility that there is some foreground or background contamination in the \TESS~light curve, due to the large size of its pixels. The {\em Gaia} eDR3 catalog \citep{2021A&A...649A...1G} shows that TIC~5724661 only has one nearby star within $80''$, and this star has $m_G$ = 19.5 -- too faint to exhibit pulsations of the amplitude that we observe. Moreover, this nearby star's {\em Gaia} BP$-$RP value is 1.74, suggesting that this is an extremely cool star that should not be able to pulsate at all \citep{andrae2018gaia}. Moreover, the Renormalized Unit Weight Error (RUWE) for TIC~5724661 is 1.482 -- which is significantly greater than the expected ``typical'' value of 1; relatively large RUWE values---usually those $\gtrsim 1.4$---can often be used as a proxy for binarity (see, e.g., \citealt{2020MNRAS.496.1922B}, and the sample selection criteria used in \citealt{2020AJ....159...19Z}). As a result, we can safely discount the possibility of contamination by another source and focus on the presence of a hot compact companion {\purple in the TIC\,5724661 system.}

We thus conclude that this high-frequency mode likely arises from a p~mode sdBV$_{\rm r}$ star. Constraints on its mass are discussed in Section~\ref{sec:spec}, its temperature in Section~\ref{sec:sed}, and its evolutionary history in Section~\ref{sec:stel-evol}.
\section{Spectral Analyses}
\label{sec:spec}
To study the spectra we obtained, we conducted two analyses---one to establish constraints on the radial velocity variations, and hence on the mass of a potential unseen sdB companion, and another in which we directly searched for spectral signatures to check the chemical composition of the A star and identify any peculiarities. 

\begin{figure}
\begin{center}
\includegraphics[width=1.0\linewidth,angle=0]{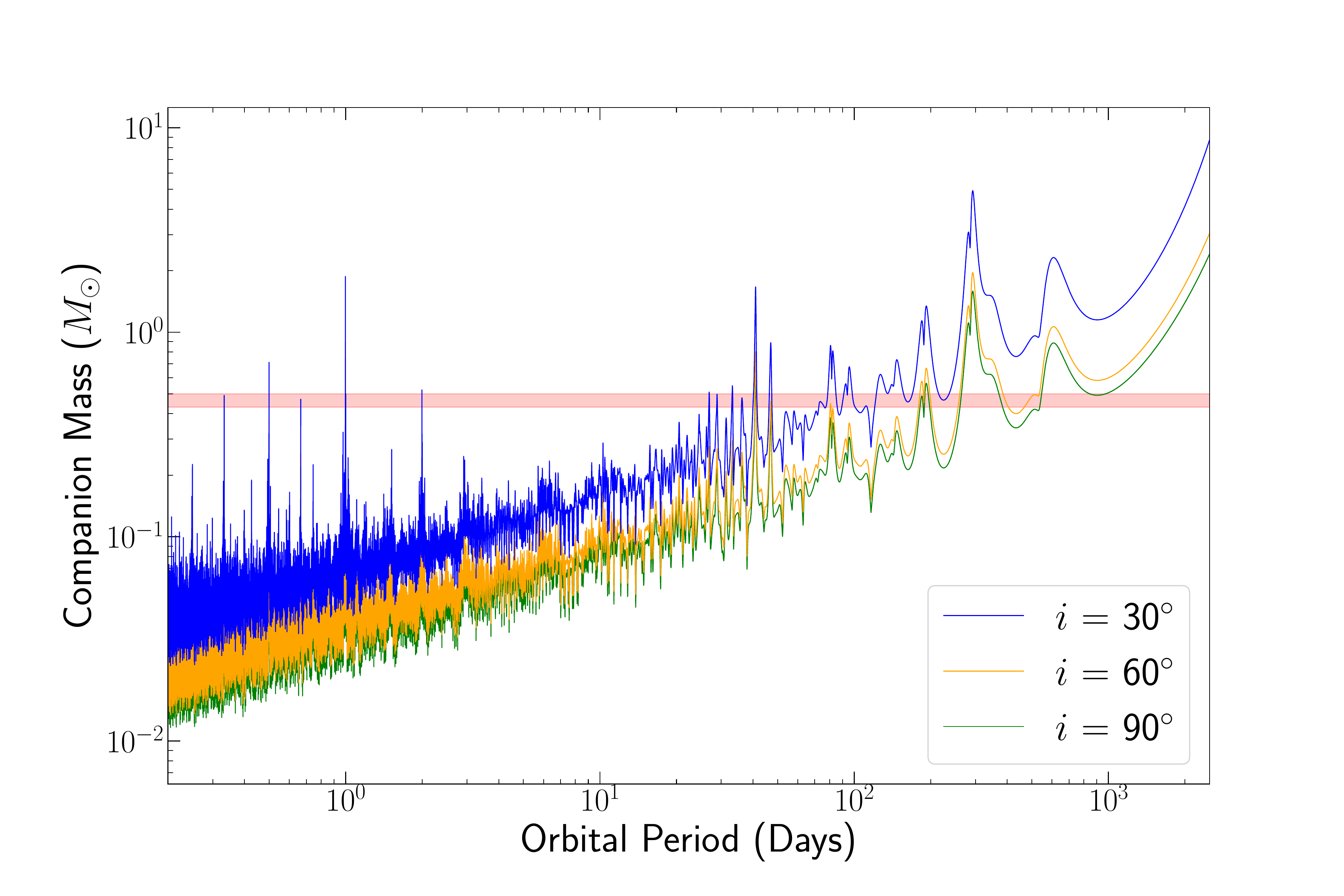}
\caption{Upper limits on the mass of a potential companion to a 2 M$_\odot$ star that are derived from our RV measurements for a range of assumed orbital inclinations. The shaded red region indicates the range of masses of the sdBs that resulted from the modeling of various evolutionary scenarios (described in Section \ref{sec:stel-evol}). It is clear that the derived constraints are more stringent for periods $\lesssim 35$~d. There exist islands of marginally acceptable binary periods between 35 and $\sim$60~d, {\purple especially for lower inclinations}; more probable periods lie near 150 and $\sim$ 300~d. {\purple The mass is essentially unconstrained above 500~d.} Spikes represent locations where we do not possess any information on the mass of a potential companion, as a result of our observing cadence. The narrow spikes below 3~d are aliases of the 1~d observing windows.}
\label{fig:companion-constraints}
\end{center}
\end{figure}

First, we used the lack of detectable RV variations to constrain the mass of a potential companion. We fit for the $K$ velocity, orbital phase, and $\gamma$ velocity of the RV curve for each of $10^6$ trial periods evenly spaced in logarithmic space between 0.1 and 1000\,d, all assuming circular orbits. To be conservative when generating our constraints, we multiplied the LSD uncertainties (described in Section\,\ref{sec:obs}) {\purple by 1.6} and input those as the argument {\texttt{sigma}} to the \texttt{curve\_fit} function in \texttt{scipy}. For each trial period, we then calculated an upper limit to the value of the mass function using the best-fit $K$ value plus twice its derived uncertainty. Finally, we solved for the corresponding limit to the mass of a potential companion by using the upper limit on the value of the mass function and an assumed mass for the A star of 2M$_\odot$. This was done for each of three assumed orbital inclinations of $30^\circ$, $60^\circ$, and $90^\circ$. A plot of the derived upper constraints for a potential companion is given in Figure \ref{fig:companion-constraints}. The data suggest that any sdB star companion is more likely to be in an orbit longer than $\sim 150$\,d. However, there are cases involving low inclination angles that could harbor either a short- or medium-period sdB star (e.g., 35--60~d). There is also the possibility the orbit is eccentric, which may lead to inauspicious locations along the orbit when the radial velocities were measured. The regions of parameter space that could result in {\purple both intermediate and long} periods {\purple for the sdB companion} is explored further in Section \ref{sec:stel-evol}.

We then turned our attention to directly analyzing the spectrum to ascertain more about the nature of the A star. We began by summing the seven TRES spectra into one, as there were no significant radial velocity differences among them. The summed spectrum had a S/N of about 65 and was, just like in Section \ref{subsec:spec}, compared to model atmospheres using ATLAS9 \citep{2004astro.ph..5087C} and SPECTRUM \citep{1994AJ....107..742G}. An atmosphere with $T_{\rm eff}=8000$~K, log~$g=4.0$, [M/H]=0, broadened to $v \sin i = 40$~km~s$^{-1}$ gave a good fit to the summed spectrum. A search for chemical peculiarities indicative of a magnetic A star yielded a null result, with the possible exception of a somewhat narrow Ca~K line at 393.366 nm. Likewise, searches for He lines in the summed spectrum caused by a possible sdB companion remained negative. This latter non-detection could be explained through a pure-H atmosphere, which may arise from chemical differentiation processes in the sdB: \citet{1981Msngr..24....7H} and \citet{2018A&A...609A..89L} suggest that processes such as gravitational settling, stellar winds (for hotter sdB/O stars), and convective instability can cause the He abundance to deviate from what is expected. On the other hand, this could simply be a consequence of an sdB companion being 2.5 magnitudes fainter than the $\delta$ Scuti star in the optical (see Fig.~\ref{fig:sed-best-fit}).

\section{Spectral Energy Distribution}
\label{sec:sed}

\begin{figure}
\begin{center}
\includegraphics[width=1.0\linewidth,angle=0]{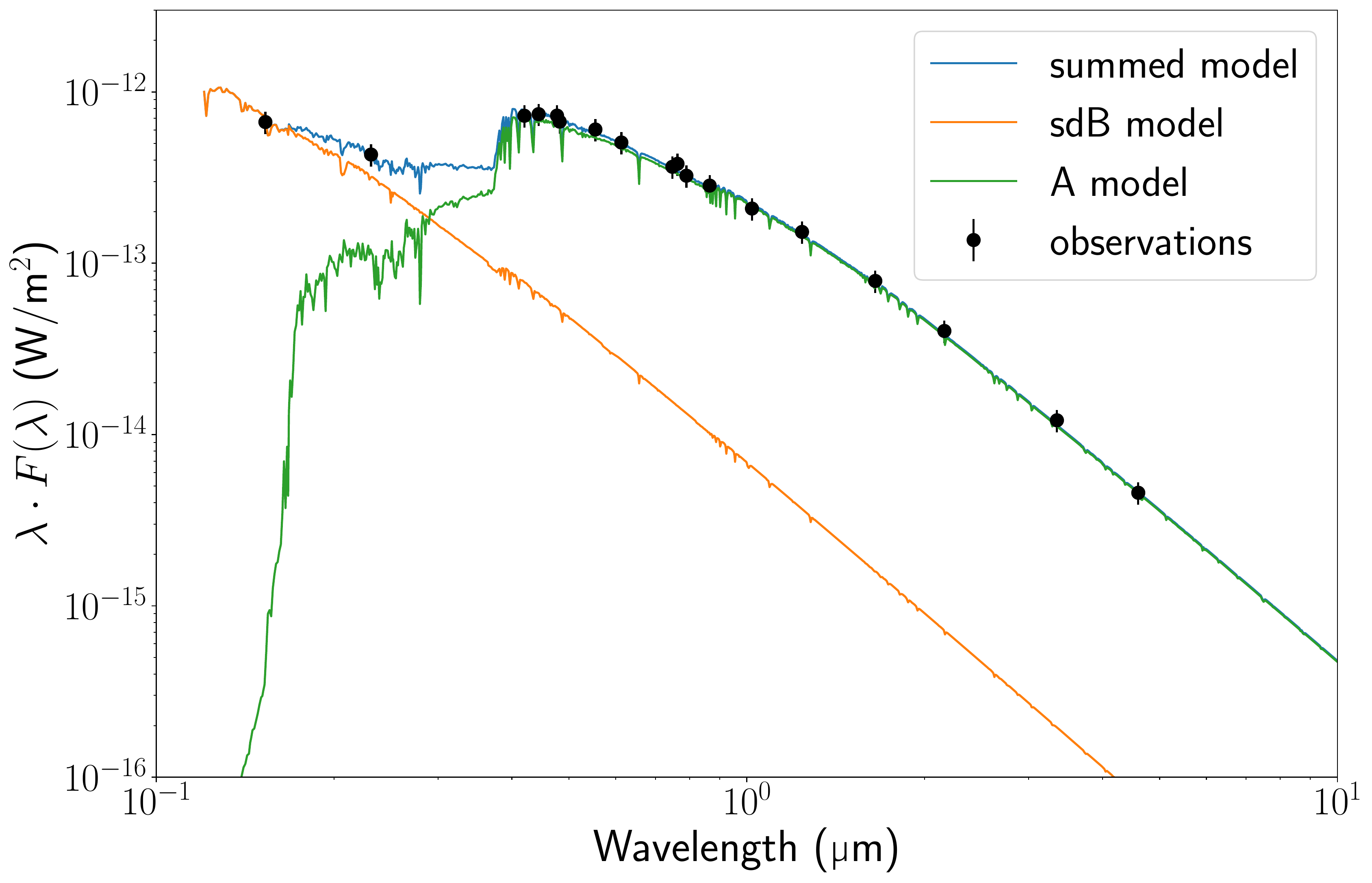}
\caption{SED plot for TIC 5724661 (black points), where the smooth curves are the model fits using Kurucz \citep{2004astro.ph..5087C} spectra for the sum of the A star and the inferred sdB star; these have been corrected for interstellar extinction. The fits are described in detail in the text. The green curve is for the A star alone, while the orange curve represents the flux of a companion sdB star. It is evident that the sum of the models for an A star and an sdB star can explain the observed SED much better than either would on their own. This lends credence to our claim that there is a compact, hot body orbiting the A star.}
\label{fig:sed-best-fit} 
\end{center}
\end{figure}

The spectral energy distribution (SED) obtained from the Vizier portal \citep{vizier} exhibits an excess in the ultraviolet flux in both the Galex NUV and FUV bands (see Fig.~\ref{fig:sed-best-fit}). Thus, we fit the SED with a model for the summed spectra from an A star and an sdB star to further test the possibility of an unresolved long-period hot sdB companion to the A star.

We used a custom implementation of the Markov Chain Monte Carlo (MCMC) algorithm to estimate parameters for the temperatures and radii of the two potential stars in the system. {\purple The extinction $A_V$ was set as a free parameter, as the estimates provided by Gaia for $A_G$ appeared to be unreliable for our purposes. Specifically, the value of $A_G$ provided in the Data Release 3 (DR3), when converted to $A_V$ using the conversion factors given in \citet{2019ApJ...877..116W}, does not agree with the value provided using the NED calculator (based on \citealt{2011ApJ...737..103S}). The extinction at other wavelengths was calculated based on the prescription given in \citet{1989ApJ...345..245C}.} The distance to the source was fixed at 713 pc, based on the {\em Gaia} parallax measurements given in DR3 \citep{gaia-mission, 2021A&A...649A...1G}. The Vizier data points were fit with summed model Kurucz spectra of an A star with fixed $\log\,g$ = 4.3 and an sdB star with fixed $\log\,g$ = 5 \citep{2004astro.ph..5087C}. {\purple While there exist significant processes in hot, compact stars that could lead to non-local thermal equilibrium (NLTE) effects in the spectral energy distributions, \citet{1994ApJ...432..351S} note that at typical surface gravities for these stars, non-LTE and LTE atmospheres agree quite well. Additionally, the key changes occur in the Balmer lines, as discussed in \citet{1997A&A...322..256N}; at the resolution of the observational data points we are using, these lines are not resolvable, making our decision to pursue an LTE analysis reasonable.} We set the priors on the A star to be 1\,R$_\odot < R_{\rm A} < 2.5$\,R$_\odot$, with $7000~{\rm K}<T_{\rm eff}<11000$~K. The sdB star's radius was sampled logarithmically and constrained to be within $0.1<R_{\rm sdB}<1$\,R$_\odot$, with $15000<T_{\rm eff}<50000$~K. 

{\purple Finally, we set the prior on $A_V$ as 0 $\leq A_V \leq 0.3$. Because we are fitting the composite spectrum of two stars in this SED, it is very helpful to have reliable prior constraints on the extinction parameter $A_V$.  The TESS input catalog v8.2 (reference), as listed on MAST\footnote{\url{https://mast.stsci.edu/portal/Mashup/Clients/Mast/Portal.html}} gives $E(B-V) = 0.021 \pm 0.005$.  For a standard conversion factor of $R_V \simeq 3.1$, this translates to $A_V \simeq 0.065$.  The NED Galactic Extinction Calculator\footnote{\url{https://ned.ipac.caltech.edu/forms/calculator.html}}, which is based on \citet{2011ApJ...737..103S}, gives $A_V \simeq 0.096$ to infinity.  However, since this source has a Galactic latitude of $-65^\circ$ and the source is 611 pc away, this is well out of the Galactic plane, and we take this to be a good representation of $A_V$ to the source itself.  HEASARC\footnote{\url{https://heasarc.gsfc.nasa.gov/docs/tools.html}} provides a hydrogen column density (also to infinity) of $N_H \simeq 2.4 \times 10^{-20}$ cm$^{-2}$.  If we adopt a conversion factor of $4.5 \times 10^{-22} \,A_V/N_H$ (as provided in \citealt{2009MNRAS.400.2050G}), we can estimate $A_V \simeq 0.11$.  Finally, we note that Gaia's early Data Release 3 \citep{2021A&A...649A...1G} lists a value for $A_G$ of 0.62.  If we converted this to $A_V$, using the relations given in Table 3 of \citet{2019ApJ...877..116W}, we would infer a value of approximately 0.79.  However, in light of the extreme disagreement with the other estimates of $A_V$, and because the Gaia estimate of $A_V$ is presumably based only on three spectral points, we discount this estimate of $A_V$ and do not use it.  Therefore, in our MCMC evaluation of the TIC\,5724661 system parameters, we set a generous prior on the range of $A_V$ to be in the range of $0.0 \leq A_V \leq 0.3$.}

To ensure confidence in our assumption that fixing the value of $\log\,g$ would not significantly impact the SED model values, we used the T\"ubingen NLTE model spectra to vary $\log\,g$ for the sdB companion. This parameter was varied from $5 \leq \log\,g \leq 6.5$ \citep{2012ascl.soft12015W}, leading to only insignificant differences in the derived SED, with the largest being a few parts per thousand of the largest SED flux value. Therefore, we were confident that we could fix the values of $\log\,g$ for both stars in the system, as described above, without losing any critical information. This assumption was borne out when we plotted the posterior distribution for this parameter, which was essentially flat---suggesting that the SED is highly insensitive to this parameter. As a result of this degeneracy, we constrained $\log\,g$ through stellar evolution modeling; see section \ref{sec:stel-evol} for more details.

\begin{figure*}
\begin{center}
\includegraphics[width=.85\linewidth,angle=0]{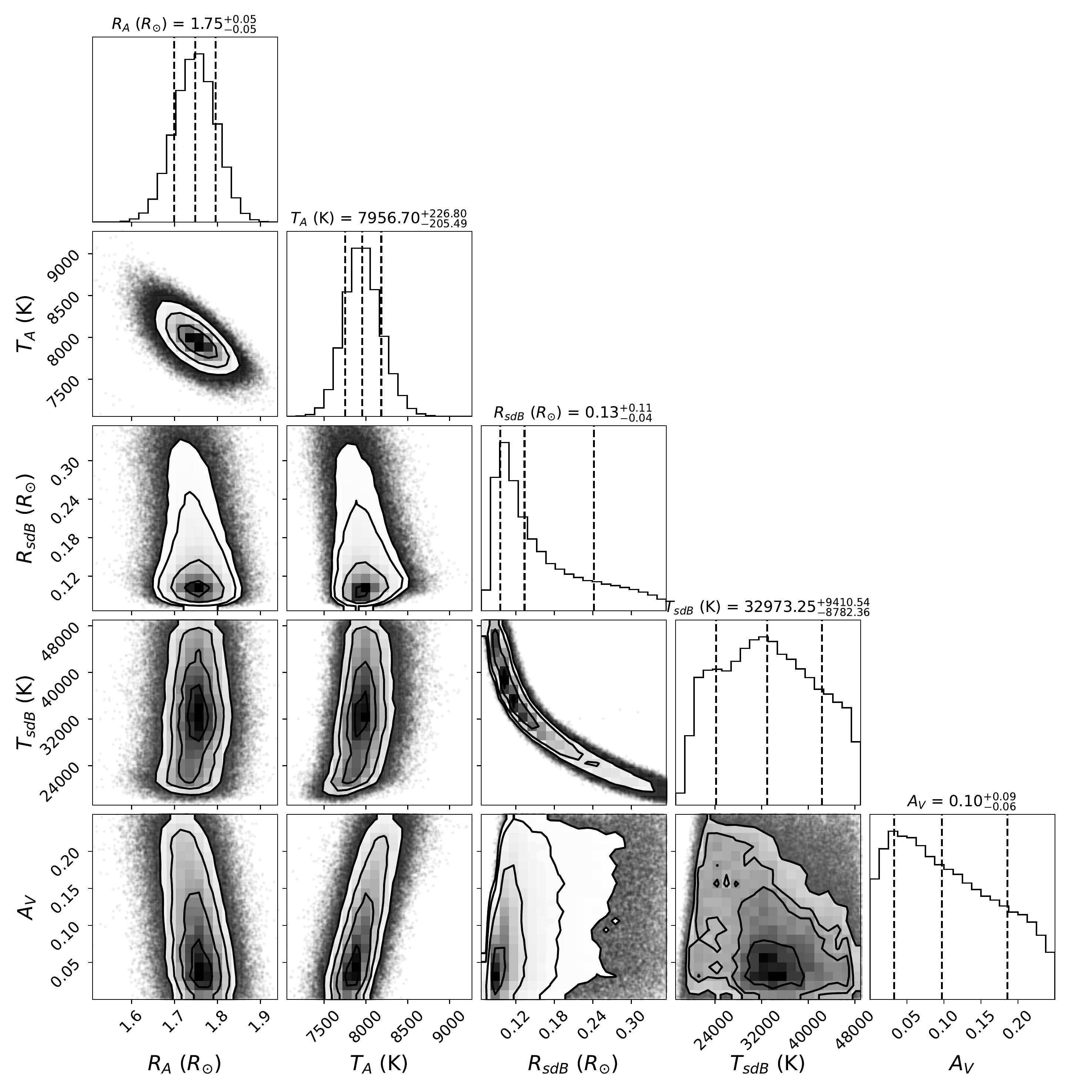}
\caption{Posterior distributions for the parameters of TIC~5724661. This corner plot shows the best-fit parameters and the correlations between the parameters derived through an MCMC fitting code. Dashed vertical lines, from left to right, represent the 16th percentile, median, and 84th percentile. The distribution for the temperature of the sdB star seems to fall off at high temperatures, suggesting that a $T_{\rm eff}$ close to 29\,000~K is most likely. There exists a strong degeneracy between the $T_{\rm eff}$ and the radius of the sdB star, as expected, given the limited region of the SED where the sdB star likely dominates the system light.}
\label{fig:sed-corner}
\end{center}
\end{figure*}

We allowed the MCMC to run for 1 million steps. The best-fit parameters for the system are presented in Table~\ref{table:mcmc-mle}, along with their associated uncertainties. Figure~\ref{fig:sed-best-fit} shows the best-fit spectrum superposed on the available data points. The fit is good, with a reduced $\chi$-squared value close to unity. A corner plot illustrating the posterior distributions and their correlations between parameters is shown in Figure~\ref{fig:sed-corner}; all parameters are somewhat correlated. There exists a strong correlation between the radius and effective temperature for the sdB star, as expected -- since its radiation dominates the observed SED only in the UV region of the spectrum. We do not show the posterior distributions for $\log\,g$, as these are flat and do not yield new information.

\begin{table}
\centering
\caption{Derived values for $T_{\rm eff}$ and $R$ for both stars.} 
\begin{tabular}{cc}
\hline
\hline
\multicolumn{1}{c}{Parameter} & \multicolumn{1}{c}{Value}
   \\
\hline
$T_{\rm eff}$ - A star & $ 7950^{+230}_{-210}$~K \\ 

$R$ - A star & $1.75 \pm 0.05$~$R_\odot$\\

$T_{\rm eff}$ - sdB star & $33000^{+9400}_{-8800}$~K\\

$R$ - sdB star & $0.13^{+0.11}_{-0.04}$~$R_\odot$\\

$A_V$ & $0.10^{+0.09}_{-0.06}$ \\
\hline
\hline
\end{tabular}
\label{table:mcmc-mle}
\end{table}

These fitted parameters for the putative sdB star agree with what is expected for the temperature of such a pulsating star. Figure 51 of \cite{heber} shows a demarcation between short- and long-period sdB pulsators, with the former having higher temperatures and $\log\,g$ values. Our results are reassuring, insofar as our inference of a pulsating sdB companion based on the observation of a {\purple high-frequency (short-period)} pulsation in the TESS data is bolstered by the value of our best-fit value for $T_{\rm eff}$ of the sdB star. However, what is unique about this sdB star is that it may lie in a little-explored region of binary parameter space: It could have an orbital period that is too long to suggest formation via common-envelope evolution, but it also could be too short to have evolved via stable {\purple mass transfer from a low-mass red giant near the tip of the red giant branch.}

\section{Evolutionary Analysis}
\label{sec:stel-evol}

Most evolutionary channels leading to the formation of hot subdwarfs rely on a red-giant progenitor that is rapidly stripped of its deep, hydrogen-rich envelope as a result of binary interactions. Once the red giant's core is exposed, it rapidly evolves along the extreme horizontal branch (EHB; see \citealt{heber}, and references therein). 

A large fraction of sdBs are found in binary systems, and the majority of these are found in short-period binaries with $P_{\rm orb} \lesssim 5$\,d {\purple (see, e.g., \citealt{2003A&A...404..301R}). Many of these have low-mass companions, such as dM or WD stars}. There clearly exists a selection effect favoring the discovery of short-period eclipsing binaries due to strong illumination effects and deep eclipses, especially for large orbital inclinations. While there exists extensive observational evidence for sdBs in short-period binaries, there have been many fewer examples of observed long-period {\purple binary systems ($P_{\rm orb} \gtrsim 300$\,d) containing sdBs (see, e.g., \citealt{2019CoSka..49..264V}, and references therein). Our analysis of TIC 5724661 suggests that its orbital period could fall in the ``intermediate'' period range, of tens to hundreds of days. If true, TIC 5724661 would fall into a sparsely populated region of parameter space and could imply a deficiency in our understanding of the formation of (binary) hot subdwarfs.} 

In this section, we discuss how to form sdBs with $P_{\rm orb} \gtrsim 70$\,d and $T_{\rm eff} \approx 30\,000$~K, as we have estimated for TIC~5724661. {\purple We analyze two types of evolutionary models---one with a low-mass progenitor ($\sim$1.2 M$_\odot$), and the other with an intermediate-mass progenitor ($\sim$3.5 M$_\odot$). These produce, respectively, long- and intermediate-period binaries containing an sdB.}

\subsection {Evolutionary Simulations}

In trying to determine the initial conditions needed to reproduce the inferred properties of TIC 5724661, we created a highly focused grid of evolutionary tracks using the {\tt MESA} binary stellar evolution code \citep{2011ApJS..192....3P,  2013ApJS..208....4P, 2015ApJS..220...15P, 2018ApJS..234...34P, 2019ApJS..243...10P}.  We had previously used {\tt MESA} to successfully explain the current evolutionary state of MWC 882 \citep{2018ApJ...854..109Z}---which itself will evolve to become an sdB---and subsequently computed a grid of about 3500 models whose initial conditions were chosen so as to optimize the likelihood of the formation of {\purple intermediate-period binary sdBs \citep{Sen2019}}.  Those models assumed varying degrees of non-conservative mass-loss and produced sdBs with a wide range of effective temperatures (20,000 K $\lesssim T_{\rm eff} \lesssim$ 50,000 K).  {\purple Using the results from this grid as our guide, we were able to optimize the computational strategy used to reproduce the properties of TIC\,5724661. In particular, we found that the best matches were obtained by assuming highly non-conservative mass transfer.}

Evolutionary tracks in this focused grid were computed using {\tt MESA} version r10108. {\purple Approximately 160 successful sdB tracks were computed}.  The sdB progenitor (i.e., the primary) was assumed to have a typical Population I metallicity ($Z = 0.02$), the atmosphere was approximated by a simple boundary condition ($\tau = 2/3$), and the local mixing-length ratio was set equal to 2. We applied the default parameters for both the Reimers' wind formula \citep{1975MSRSL...8..369R} and the Bl\"ocker wind formula \citep{1995A&A...299..755B}. We tested a reasonable range of other values for these parameters and found that they had a {\purple small} effect on the results. The most important factors influencing the evolution, other than $M_{1,0}$, $M_{2,0}$, and $P_{\rm orb, 0}$ (i.e., the initial mass of the primary, the initial mass of the secondary, and the initial orbital period, respectively), were the parameters $\alpha$ and $\beta$ \citep{2006csxs.book..623T}. The parameter $\alpha$ is the fraction of the mass lost from the primary (donor) and then {\purple directly} ejected from the binary, carrying away the specific angular momentum of the primary.  Similarly, $\beta$ is the fraction of the mass transferred from the primary (donor) to the secondary (accretor) that is subsequently lost from the binary, carrying away the specific angular momentum of the secondary.{\footnote{Both cases correspond to the ``fast Jeans' mode'' of angular momentum dissipation.}}$^{,}${\footnote {$\beta$ can equivalently be viewed as the fraction of mass lost from the binary \textit{after} it has crossed the L1 point.}}  {\purple We can express the amount of mass that has been accreted by the secondary as 
\begin{equation}
\delta M_{2}=-(1-\alpha -\beta)\delta M_{1}\qquad 
\end{equation}
To simplify the analysis, we eliminated one extra dimension of parameter space in our computations by setting $\alpha =0$.} The main justification for this strategy is our (empirical) finding that the value of $\alpha + \beta$ had a much greater impact on the evolution than did various combinations of those parameters corresponding to the same sum. Our choice of $\beta$ determined the degree to which mass transfer was non-conservative.

\begin{figure}
\begin{center}
\includegraphics[width=1.0\linewidth,angle=0]{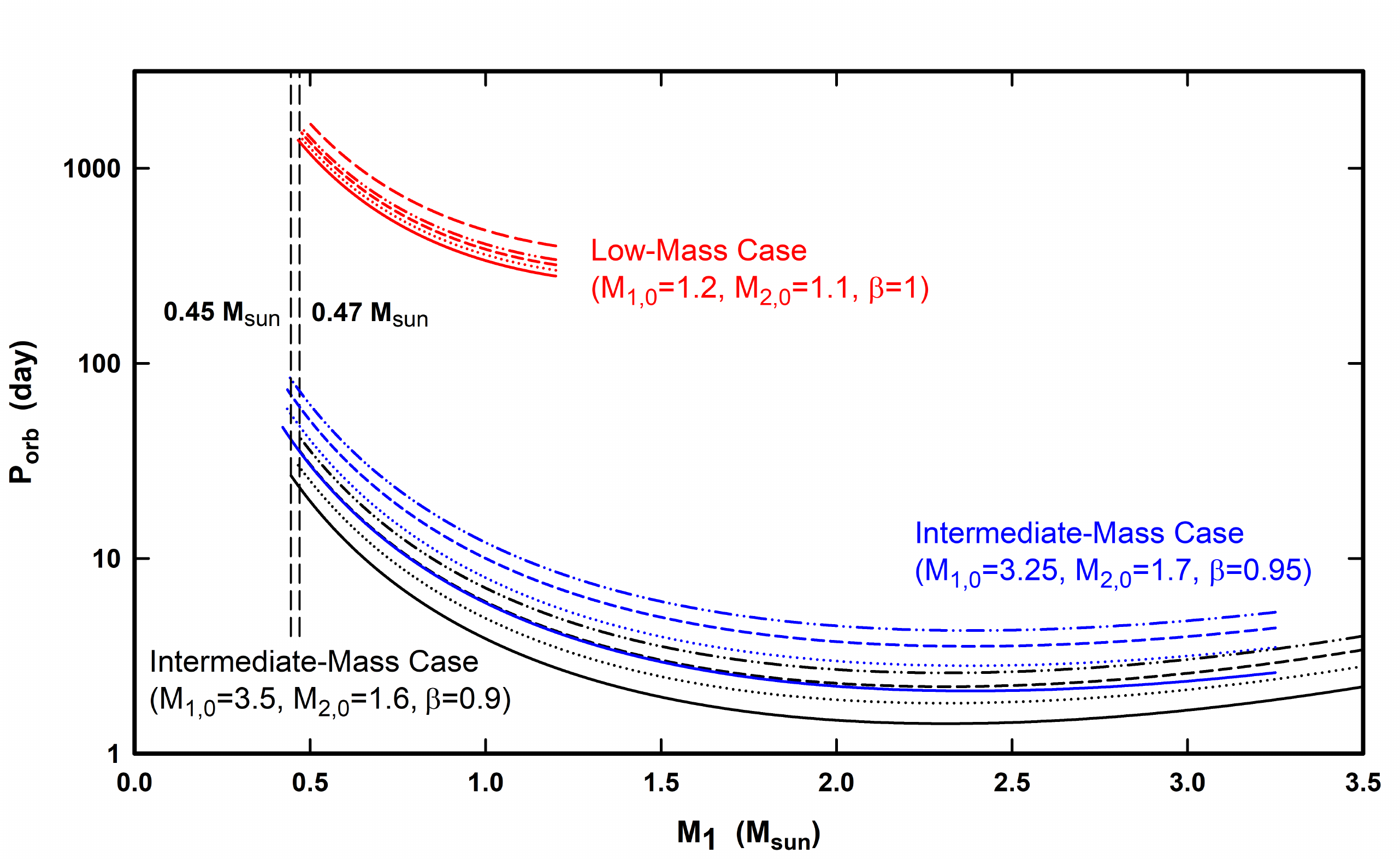} 
\caption{{\purple The evolution of the orbital period as a function of the mass of the primary (i.e., the sdB star's progenitor). Representative binaries for both the intermediate- and low-mass cases are shown. The respective initial masses of the primary and the secondary (in solar units) and the value of $\beta$ for each of the three sets of curves are listed in the diagram. For the black curves, the initial orbital periods are 2.2, 2.8, 3.4, and 4.0 days (solid, dotted, dashed, and dash-dotted lines, respectively) and for the blue curves, the initial periods are 2.6, 3.5, 4.4, and 5.3 days (solid, dotted, dashed, dash-dotted lines, respectively). For the low-mass cases (red curves), the initial orbital periods are 280, 300, 320, 340, and 400 days (solid, dotted, dashed, dash-dotted, and long-dashed lines, respectively). The ``canonical'' range of masses for sdB stars (0.45 -- 0.47 M$_\odot $) is denoted by the two vertical dashed lines.}}
\label{fig:pvsm} 
\end{center}
\end{figure} 

\begin{figure}
    \centering
    \includegraphics[width=\linewidth]{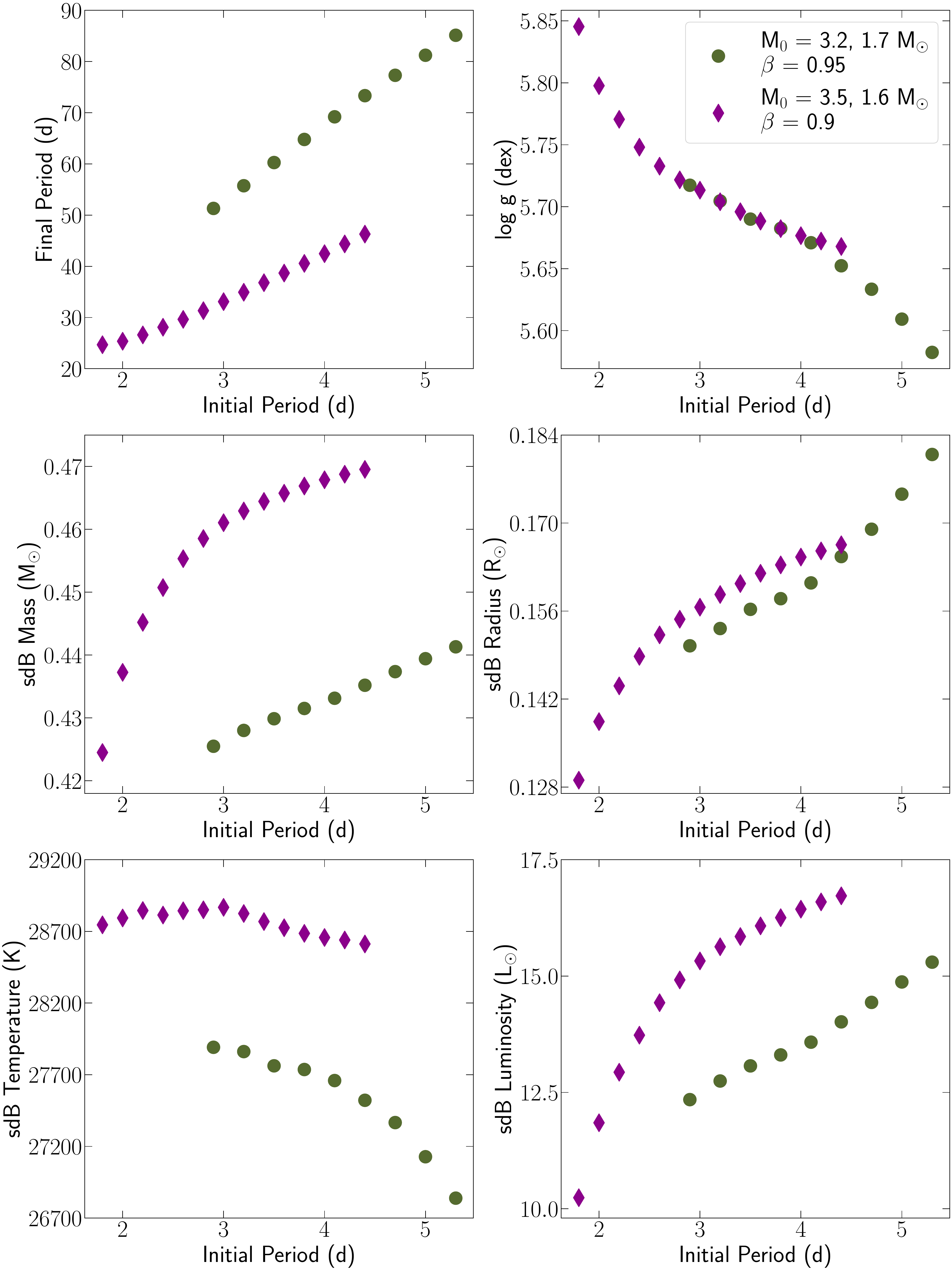}
    \caption{A multi-panel plot showing how various properties of the sdB components of two representative {\purple binaries} correlate with their respective initial periods, for {\purple two} intermediate-mass cases. The dark green dots correspond to a system in which the initial mass of the sdB star's progenitor was 3.25 M$_\odot$, its companion's initial mass was 1.7 M$_\odot$, and $\beta$ was fixed at 0.95. {\purple The purple diamonds correspond to a system in which the initial mass of the sdB star's progenitor was 3.5 M$_\odot$, its companion's initial mass was 1.6 M$_\odot$, and $\beta$ was fixed at 0.9.} The plots show that for larger initial periods, the sdB star's final period, mass, radius, and luminosity increase; however, its effective temperature and $\log g$ decrease. {\purple These trends remain robust even when the initial masses are changed. Note that we have employed a time-average for all parameters whose values change (e.g., the luminosity) during the sdB phase.}}
\label{fig:pinit_plot} 
\end{figure}

Both binary stars are evolved contemporaneously with {\tt MESA}. It is important to follow the evolution of the secondary as it accretes mass, as the secondary could expand to fill its Roche lobe.{\footnote{If the primary is still transferring mass, the resulting evolution might lead to a merger.}} The reasons why the secondary can potentially fill its Roche lobe are as follows:  (1) if the mass accretion rate onto the secondary ($\dot M_2$) is too high {\purple (i.e., the mass accretion timescale is shorter than the Kelvin time), the accretor} can expand adiabatically {\purple if it has a convective envelope}; or, (2) if the mass of the secondary were to increase substantially on a short timescale, then it could evolve to become a giant (and fill its Roche lobe) before the primary (donor) has had a chance to complete its sdB phase.{\footnote{Moreover, with respect to TIC~5724661, this would be especially problematic because the giant would be more luminous than the sdB star (contrary to observations), and it would not exhibit $\delta$-Scuti-like pulsations.}} In either case, {\tt MESA} halts further computation. In order to increase the chances that the primary evolves through the sdB phase, we typically attenuated the mass accretion rate onto the secondary by requiring that $\beta \gtrsim $ 0.8 (recall that $\dot M_2 = -(1 - \beta) \dot M_1$). Obviously, if mass transfer is fully non-conservative ($\beta = 1$), the secondary is not likely to fill its Roche lobe until long after the sdB phase is complete {\purple (assuming the primary evolves along the EHB).

Because of the potential importance of the evolution of wide sdB binaries in explaining the properties of TIC ~5724661, we have also computed the evolution of a small grid (about 50 models) of primordial binaries that are composed of low-mass stars (M$_{1,0}$ = 1.2\,M$_\odot$) for several initial orbital periods and primordial mass ratios. \citet{2020A&A...641A.163V} showed that if the primary (donor) can evolve up the red giant branch and mass transfer is initiated via stable Roche lobe overflow close to the point of helium flash (i.e., the tip of the red giant branch), then the giant primary can be stripped of its hydrogen-rich mass on a very short timescale, leaving behind a remnant core that subsequently undergoes helium burning and leads to an sdB phase. Unlike the evolution of intermediate-mass stars discussed above, this low-mass channel produces very wide sdB binaries. The large radius of the giant combined with the constraints imposed by Roche lobe geometry enforce a wide separation at the onset of mass transfer. This separation becomes wider as the binary evolves through the mass-transfer phase. Intermediate-mass stars, on the other hand, typically initiate mass transfer at much smaller separations (and shorter orbital periods) because they do not need to be as highly evolved at the onset of mass transfer in order to achieve helium burning in the stripped core. As mass transfer proceeds their separations also widen. The evolution of the orbital period as a function of the decreasing mass of the primary (i.e., the sdB star's progenitor) is shown for representative cases in Figure \ref{fig:pvsm}.}

\subsection {Results}

\begin{figure*}
\begin{center}
\includegraphics[width=.95\linewidth,angle=0]{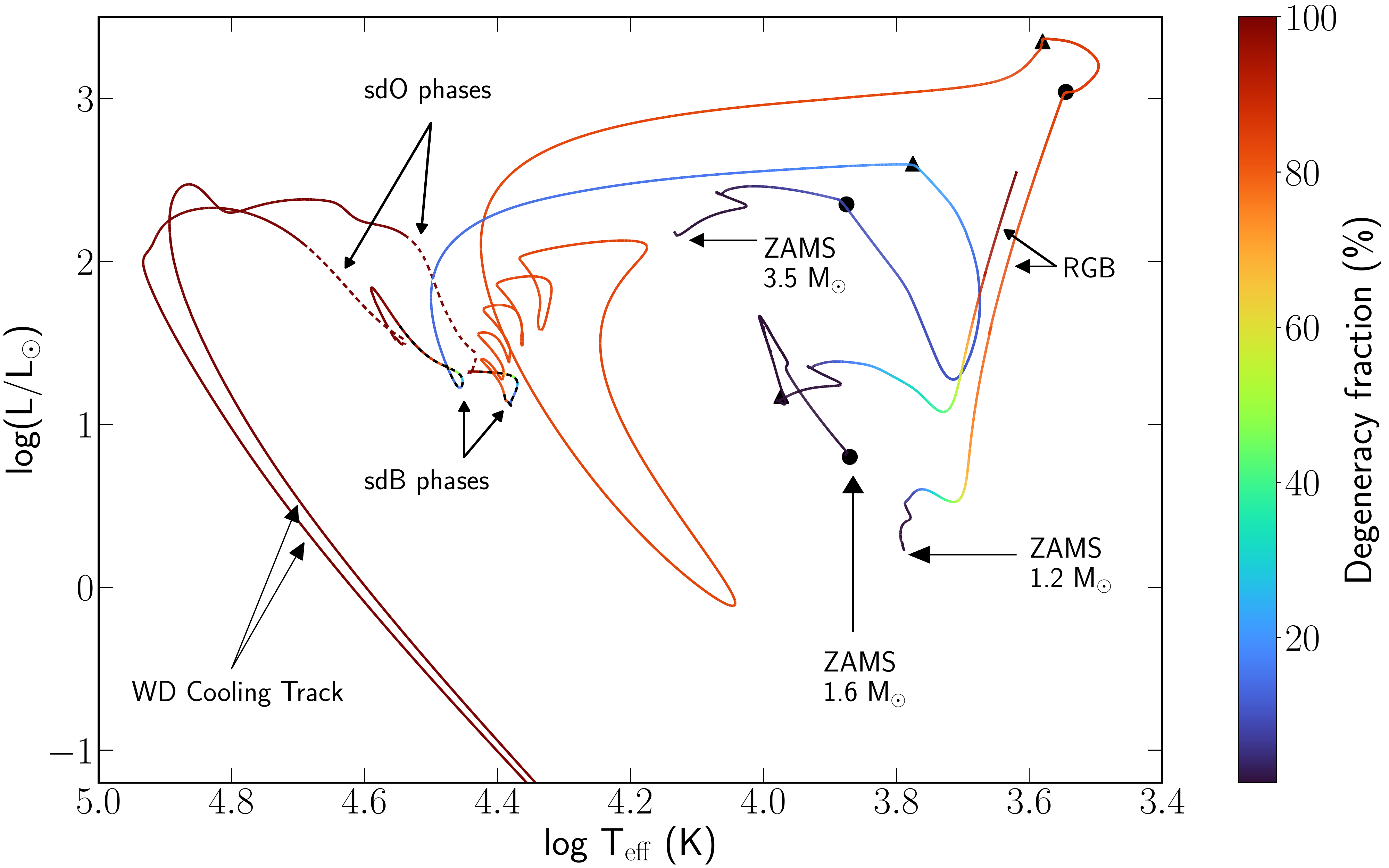}
\caption{Formation and evolution of a representative intermediate-period and long-period sdB from the zero-age main sequence (ZAMS) to the final WD cooling stage in the H-R diagram. {\purple The solid lines denote the evolutionary tracks and the corresponding colors (as given by the color bar) indicate the percentage of the combined ideal gas and electron degeneracy pressures that is solely due to degeneracy (evaluated at the center). A filled circle denotes the start of a mass-transfer phase and a triangle denotes the end of that phase.} The ZAMS progenitor star for the intermediate-period case has a mass of 3.5 M$_\odot$ and an approximate solar metallicity ($Z=0.02$).  Mass transfer is initiated in the Hertzsprung Gap and continues as the progenitor ascends the Red Giant Branch (denoted as RGB).  Mass transfer ceases once its mass is reduced to 0.47 M$_\odot$ and this stripped core evolves along the horizontal branch until it reaches the sdB phase. This phase {\purple (annotated and denoted by the black dashed lines overlaid on the evolution curves)} persists for almost 80 Myr before the hot subdwarf {\purple subsequently} evolves through the sdO phase {\purple (distinguished by the brown dashed lines)} for an additional 40 Myr.  The evolution of the companion star is also shown; its initial mass is 1.6 M$_\odot$, and it undergoes a phase of rapid accretion before it reaches thermal equilibrium (after mass transfer has ceased) and evolves normally as a 1.9 M$_\odot$ star.  The star ascends the RGB, and the evolution is halted once the star is large enough to fill its Roche lobe. {\purple The progenitor star of the long-period sdB binary has a primordial mass of 1.2 M$_\odot$ and evolves up the RGB until it is close to its tip, at which point it starts to lose mass rapidly.  After a series of thermal adjustments, it enters the sdB phase, which persists for $\approx 75$ Myr (annotated); for reasons of clarity, the evolution of its companion is not shown. Note that the observable properties of the sdBs in both the intermediate- and long-period scenarios are quite similar, as are the fractions of electron degeneracy pressure in both stars during most of their sdB and sdO phases. }}
\label{fig:evol} 
\end{center}
\end{figure*}

\begin{figure}
\begin{center}
\includegraphics[width=1.0\linewidth,angle=0]{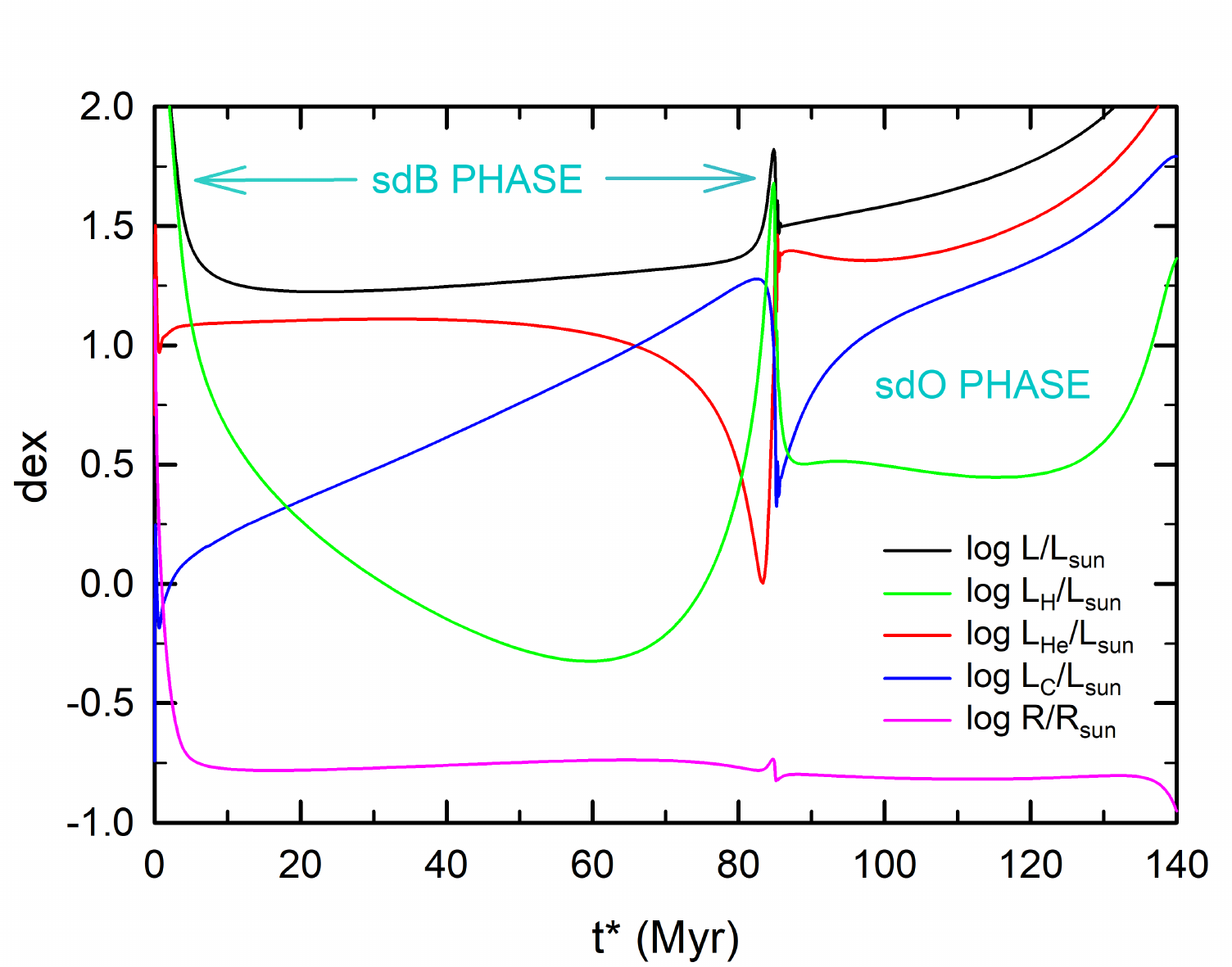}
\caption{Evolution of the bolometric luminosity, individual nuclear luminosities, and radius of the representative hot subdwarf as a function of time. Note that $t^* = 0$ has been chosen to approximately coincide with the end of the RGB phase ($t^* = t - 211.25$ Myr). The sdB phase persists for nearly 80 Myr, during which time helium is continuously depleted due to triple-$\alpha$ burning and the fusion of carbon into oxygen ($\alpha$-channel burning). Once a CO core forms and the subdwarf experiences He-shell flashes, it enters into a shorter-lived sdO phase. For each phase, the radius of the subdwarf is approximately constant.}
\label{fig:lum_rad} 
\end{center}
\end{figure} 

Using the methods described in the previous subsection, we show that the inferred observational parameters for TIC~5724661 are reproducible as long as we are willing to {\purple allow mass transfer to be highly non-conservative. We will first consider the evolution of intermediate-mass primordial binaries and show that they can produce sdBs in binaries with intermediate orbital periods. We then discuss the evolution of low-mass primordial binaries and show that they can evolve to become long-period sdB binaries.  Finally, we compare the properties of the sdBs predicted by these two channels and discuss the implications for TIC~5724661.

Based on a grid of over 3000 models from \citet{Sen2019} whose resolution was subsequently refined for TIC\,5724661, we found that comparable sdB models with intermediate periods could be obtained from a population of primordial binaries with $M_{1,0} \approx 3.5 \pm 0.3$ M$_\odot$, $M_{2,0} \approx 1.6 \pm 0.3$ M$_\odot$, $P_{\rm orb,0} \approx 4 \pm 2$\,d, and $\beta \gtrsim 0.8$. For these initial conditions (and assuming a solar metallicity), we were able to produce multiple tracks for which the sdB's effective temperature was between $\approx$27,000 --32,000 K, its $\log g$ between 5.4--5.8, and its final orbital period in the tens of days.\footnote{Longer-period sdB binary models ($P_{\rm orb} \sim 1000$\,d) are discussed later in this section.} These results are in general agreement with the inferred stellar parameters of the components of the TIC 5724661 system} (Table \ref{table:mcmc-mle}). 

Figure \ref{fig:pinit_plot} provides an example of how the predicted properties of the sdB depend on one of the dimensions of initial parameter space (i.e., $P_{\rm orb,0}$).  For {\purple the two representative cases shown in this figure,} the initial conditions for the primordial binaries were $M_{1,0} = 3.25$\,M$_\odot$, $M_{2,0} = 1.7$\,M$_\odot$, with $\beta = 0.95$, {\purple and $M_{1,0} = 3.5$\,M$_\odot$, $M_{2,0} = 1.6$\,M$_\odot$, with $\beta = 0.9$}. Not surprisingly, increasing the initial period serves to monotonically increase the final period.  Increasing the initial period implies that the donor star (primary) will be more evolved at the onset of mass transfer. This also implies an increased mass, radius, and luminosity of the resulting sdB star, but tends to lower its effective temperature.

The HR diagram for one of our representative cases that very closely reproduces the properties of the sdB in TIC 5724661 is shown in Figure \ref{fig:evol}. The sdB progenitor has a mass of 3.5 M$_\odot$, with a Population I metallicity of $Z = 0.02$; the value of $\beta$ was set equal to 0.9.  Mass transfer commences in the Hertzsprung Gap and continues as the progenitor ascends the Red Giant Branch.  Mass transfer rates from the donor sometimes exceeded 10$^{-6}$ M$_\odot$/yr, resulting in a large fraction of the giant donor's hydrogen-rich envelope being lost rapidly to the interstellar medium. Mass transfer ceases once the giant's highly mass-depleted envelope collapses, {\purple causing the star to lose contact with its Roche lobe. At this juncture,} the primary is essentially a hot helium core of mass 0.470 M$_\odot$, and it contracts rapidly (in $\sim\,3 \times 10^5$ yr) along the horizontal branch before entering a long-lived sdB phase ($\approx$ 80 Myr); the sdB phase is appropriately annotated in the evolutionary track in Figure \ref{fig:evol}.  

{\purple One of the hallmarks of the sdB phase is the relative constancy of the sdB's radius for $\gtrsim 10$ Myr (see Figure \ref{fig:lum_rad}).  We define this phase to extend from the point at which: (1) the radius has contracted sufficiently so as to remain approximately constant (for at least millions of years); and, (2) the star has increased its central carbon mass fraction by at least 1\% above its primordial value due to helium burning. Both of these conditions must be met. The sdB phase persists up} to the point when a (convective) CO-rich core first emerges. It is at this juncture that late-stage thermonuclear flashes can occur and persist briefly before the hot subdwarf enters the sdO phase. During this stage, the CO core can grow substantially in mass as the result of He-burning in a shell surrounding the core. Once the CO core has grown to reach about ~95\% of the total mass, the radius of the hot subdwarf contracts rapidly---signaling the termination of the sdO phase. Subsequently, the thin H-rich layer ($\sim 0.003$M$_\odot$) near the surface can be compressed and concomitantly heated as a result of the envelope's rapid contraction. This temperature increase is often significant enough for the star to undergo one or more shell flashes (see, e.g., \citealt{2004ApJ...616.1124N}, and references therein). Once all nuclear burning is quenched, the star descends onto the white dwarf cooling track.

Another curve in Figure \ref{fig:evol} shows the evolution of the secondary star (i.e., the accretor).  Its initial mass is 1.6 M$_\odot$, and it undergoes a phase of rapid accretion before reaching thermal equilibrium (after mass accretion has ceased) {\purple as a 1.9 M$_\odot$ MS star.  Because the accretion timescale onto the secondary is similar in magnitude to its thermal timescale, the accretor can adjust its internal structure on the order of several million years.  Its nuclear timescale is much longer (by more than two orders of magnitude), so it takes $>$0.5 Gyr for the star to evolve off of the MS; by then, the sdB is already evolving on the WD cooling track. Thus, the secondary will be observed as a MS star during both the sdB and sdO phases of evolution; the secondary then evolves into a subgiant before ascending the RGB.} \texttt{MESA} halts the evolution once the secondary fills its Roche lobe (corresponding to the end point seen in Figure \ref{fig:evol}). If we continued to follow the evolution of this binary, we would see a subsequent phase of common-envelope evolution, resulting in the formation of a double-degenerate binary. The expected end product would thus be a 0.47 M$_\odot$ white dwarf (the sdB component) in close orbit with a lower mass helium white dwarf (the core of the giant secondary). 

{\purple A third line in Figure \ref{fig:evol} shows the evolution of a representative low-mass progenitor star that produces an sdB in a wide orbit.  The primordial binary consists of primary and secondary stars with masses of 1.2 and 1.1 M$_\odot$, respectively. Consistent with the models discussed in \citet{2020A&A...641A.163V}, the primary (donor) evolves up the RGB until it is close to the tip of the RGB (where a helium flash is expected to occur). The star then starts to lose mass very rapidly via Roche lobe overflow, with a mass-loss timescale of $\sim 10^{6}$ yr. After a series of flashes, it enters the sdB phase (as defined previously), which persists for $\approx 75$ Myr. The star then evolves through the sdO phase (annotated on the plot), before eventually cooling as a WD. 

So what are the similarities and differences between the properties of the sdB binaries formed by these two channels?  The masses of the two sdBs are virtually identical, and are close to the canonical mass for such stars (0.470 M$_\odot$ compared to 0.466 M$_\odot$).  As seen in the HRD, both have very similar luminosities, effective temperatures, and thus radii.  Their (time-averaged) central temperatures and densities (and thus fractional electron degeneracy pressures) are very similar during their respective sdB phases.  Their sdB lifetimes persist for nearly 80 Myr (each), and even their hydrogen-rich envelopes have the same mass to within a factor of 2 ($\approx$0.0025 M$_\odot$ compared to $\approx$0.005 M$_\odot$). One difference relates to the percentage of electron degeneracy pressure at the center of the primary star preceding the sdB phase: The primary of the low-mass case has a largely degenerate core, while the intermediate-mass primary is only partially degenerate. However, soon after the start of the sdB phase, both are partially degenerate ($\approx 20$\% of the central pressure is due to electron degeneracy), and they evolve to become fully degenerate by the end of the sdB phase. The major difference between the two models, as expected, is in $P_{\rm orb}$. The intermediate-mass channel produced an sdB binary with $P_{\rm orb} \simeq$46 d, while the low-mass channel produced an sdB binary with $P_{\rm orb} \simeq$1395 d.  Although there is a range of possible orbital periods according to both scenarios, it is fair to say that the low-mass case can produce a binary sdB that has a period approximately an order of magnitude larger than that of the intermediate-mass case. 

Figure \ref{fig:lum_rad} provides a more detailed perspective with respect to the sdB and sdO evolutionary phases for the intermediate-mass case.} It shows the temporal evolution of the nuclear luminosities from the H-, He-, and C-burning channels, and the evolution of the surface luminosity{\footnote{Note that the sum of the nuclear luminosities may not equate to the surface luminosity. This difference is due to the gravothermal luminosity, whose magnitude is not shown.}} and radius. The high luminosities seen near $t^* = 0$ ($\equiv t - 211.25$ Myr) arise from the evolution of the primary star while it is still a red giant. After the star settles into the sdB phase, its radius remains approximately constant for $\approx$78 Myr. Initially, He-burning accounts for most of the luminosity, but as more and more carbon is created, $\alpha$-channel capture occurs, converting some of the carbon into oxygen. As both the luminosity and radius approach a local maximum, a convective core ($\approx 0.1$ M$_\odot $) of CO is formed, and the hot subdwarf thermally relaxes, leading to a brief phase of shell flashes. The subdwarf subsequently enters the sdO phase, during which time the radius is reasonably constant over $\approx 40$ Myr. Theoretically speaking, we can think of the sdB and sdO phases as being long-lasting ($>$ 10 Myr) and quasi-quiescent. The hallmark of the sdB phase is He-burning in the core; however, for the sdO phase, He-burning mainly occurs in a shell around the CO core.

{\purple For the intermediate-mass case, our models show that the mass of the A star companion (i.e., the secondary) could lie in the range of $\approx$1.8 to 1.9 M$_\odot$.}  The main reason that the secondary masses are so large is that: (i) the intial mass ratio ($M_{2,0} / M_{1,0}$) must be sufficiently high at the onset of mass transfer ($\gtrsim 0.4$) to avoid dynamical instability (otherwise, this could lead to a merger); and, (ii) $M_{1,0}$ must be $\gtrsim 3$M$_\odot$ so that the primary has a chance to initiate core He-burning after departing the red giant branch.{\footnote {If $M_{\rm 1,0} \lesssim 3$M$_\odot$, the primary will evolve into a helium white dwarf and never undergo an sdB phase for the range of initial conditions that we considered.}}  According to the main sequence models of \citealt{Eker_2015}, a 1.9-M$_\odot$ solar-metallicity model has a luminosity of about 15 L$_\odot$, $T_{\rm eff} \approx$\,8000\,K, and $R \approx 2$ R$_\odot$. Its inferred spectral type is A2.5V.  These values are in line with the inferred values shown in Table \ref{table:mcmc-mle}. {\purple For the low-mass case, it appears that the constraints on the secondary's mass are less restrictive, but we would need to compute many more of these types of models before coming to a definitive conclusion.}

\section{Conclusions}  
In this paper, we present strong evidence for the nature of TIC~5724661: a main-sequence A star with an inferred long-period sdBV$_{\rm r}$ star in orbit around it. First, we used radial velocity data to show that the putative hot subdwarf companion must have a period longer than a few tens of days. We then fit the spectral energy distribution using an MCMC code to constrain the parameters of the two stars in the system and provide fairly compelling evidence that this star is indeed an sdB star. To determine whether such sdBs can be produced using non-common-envelope formation channels, we modeled this system with MESA and demonstrated that we can readily produce such stars {\purple with either an intermediate- or a low-mass progenitor}, as long as there is a high degree of non-conservative mass transfer. We expect that the current A star in TIC~5724661 could have accreted perhaps $\sim 10\%$ of the mass lost from the sdB's progenitor, implying a final mass of 1.9 M$_\odot$. This would ensure that this star, which exhibits $\delta$-Scuti pulsations, is an A star (A2.5V). {\purple A more limiting constraint on the minimum value of $P_{\rm orb}$, perhaps via more spectroscopic observations, may serve to either strengthen or invalidate either of the intermediate-mass or low-mass progenitor binary models that we have proposed in Section \ref{sec:stel-evol}}. 

Our work adds more observational evidence for intermediate- and long-period binaries that contain an sdB component (for earlier examples of such binaries, see, e.g., \citealt{2019CoSka..49..264V, 2020A&A...641A.163V}, and the references therein). Previously, many observed sdB binaries were found with extremely short periods and were therefore thought to have passed through a common-envelope phase. Our work lends observational evidence to the fact that such common envelopes are not at all necessary to form sdB stars (see, e.g., \citealt{Sen2019} for a set of intermediate-period sdB formation models), especially those found to pulsate in high-frequency p~modes.

Finally, this paper---which utilizes the TESS 20-second cadence data---further emphasizes the power of TESS for precision asteroseismology. The continuous, short-cadence nature of TESS data enables the detection and study of objects and pulsations that ground-based campaigns have, in the past, struggled to identify and characterize.\footnote{{\purple For an example of a ground-based photometric campaign to study high-frequency pulsations, such as those in TIC\,5724661, see, e.g., the Whole Earth Telescope \citep{1990ApJ...361..309N}}.} As a result, we were able to identify and derive a precise frequency estimate for the sdB pulsation frequency and, consequently, recognize the presence of a companion. Future work on this object will focus on long-term radial velocity monitoring of this star in order to better constrain the true period of this binary, {\purple as well as further modeling to break the degeneracy between the intermediate- and long-period cases}.

\section{Acknowledgments}

{\purple We thank the anonymous referee for pointing out several issues that motivated us to carry out additional analyses and clarify some of our discussions. We acknowledge Andrzej Baran, Michael Fausnaugh, and Jon Jenkins for helpful discussions regarding the TESS data and the locations of the pulsations.} We also thank St\'ephane Charpinet and JJ Hermes for helping us interpret our estimates for the $T_{\rm eff}$ of the sdB star. L.\,N.~thanks the Natural Sciences and Engineering Research Council (Canada) for financial support through the Discovery Grants program. G.\,H.~acknowledges support by the Polish NCN grants 2015/18/A/ST9/00578 and 2021/43/B/ST9/02972.

This paper includes data collected by the TESS mission, specifically through Director's Discretionary Target programs \#22 (PI: Antoci) and \#46 (PI: Jayaraman). Funding for TESS is provided by NASA's Science Mission Directorate. {\purple Resources used in this work were provided by the NASA High-End Computing (HEC) Program through the NASA Advanced Supercomputing (NAS) Division at Ames Research Center for the production of the SPOC data products.} Some computations were carried out on the supercomputers managed by Calcul Qu\'ebec and Compute Canada. The operation of these supercomputers is funded by the Canada Foundation for Innovation (CFI), and the Fonds de recherche du Qu\'ebec -- Nature et technologies (FRQNT).

This work has used data from the European Space Agency (ESA) mission {\em Gaia} (\url{https://www.cosmos.esa.int/gaia}), processed by the {\em Gaia} Data Processing and Analysis Consortium (DPAC, \url{https://www.cosmos.esa.int/web/gaia/dpac/consortium}). Funding for DPAC has been provided by national institutions, in particular those party to the Gaia Multilateral Agreement. 

We would like to acknowledge the Indigenous Peoples as the traditional stewards of the land on which part of this research was conducted, and the enduring relationship between them and their traditional territories. We acknowledge the painful history of genocide and forced occupation of their territory, and we honor and respect the many diverse Indigenous people connected to this land on which we research and gather.

{\purple Code and inlists used for our MESA analysis are available on Zenodo, at this\,\dataset[link]{https://zenodo.org/record/6668950\#.YrLI5HbMKUk}.} 

\facilities{TESS, Gaia, FLWO:1.5m}

\software{SPOC \citep{jenkinsSPOC2016}, {\tt numpy} \citep{harris2020array}, {\tt matplotlib} \citep{Hunter:2007}, 
{\tt scipy} \citep{2020SciPy-NMeth}, {\tt astropy} \citep{astropy:2013, astropy:2018}, {\tt pysynphot} \citep{2013ascl.soft03023S}, 
{\tt pandas} \citep{reback2020pandas, mckinney-proc-scipy-2010}}, {\tt SPECTRUM} \citep{1999ascl.soft10002G}, {\tt MESA} \citep{2011ApJS..192....3P,  2013ApJS..208....4P, 2015ApJS..220...15P, 2018ApJS..234...34P, 2019ApJS..243...10P}

\bibliography{5724661}{}
\bibliographystyle{aasjournal}
\end{document}